# Combating small molecule aggregation with machine learning

Kuan Lee[1], Ann Yang[1], Yen-Chu Lin[1,*], Daniel Reker[2], Gonçalo J. L. Bernardes[3,4], Tiago Rodrigues[4,5,*]

1. Insilico Medicine Taiwan Ltd., Keelung Rd., Xinyi Dist., Taipei, Taiwan
2. Department of Biomedical Engineering, Duke University, Durham, NC 27708, USA.
3. Department of Chemistry, University of Cambridge, Lensfield Road, CB2 1EW, Cambridge, UK.
4. Instituto de Medicina Molecular, Faculdade de Medicina da Universidade de Lisboa, Av. Prof. Egas Moniz 1649-028 Lisboa, Portugal.
5. Instituto de Investigação do Medicamento (iMed.ULisboa), Faculdade de Farmácia, Universidade de Lisboa, Av. Prof. Gama Pinto, 1649-003 Lisboa, Portugal.

* Corresponding authors: jimmy.lin@insilicomedicine.com
tiago.rodrigues@ff.ulisboa.pt

**Abstract**

Biological screens are plagued by false positive hits resulting from aggregation. Thus, methods to triage small colloidally aggregating molecules (SCAMs) are in high demand. Herein, we disclose a bespoke machine-learning tool to confidently and intelligibly flag such entities. Our data demonstrate an unprecedented utility of machine learning for predicting SCAMs, achieving 80% of correct predictions in a challenging out-of-sample validation. The tool outperformed a panel of expert chemists, who correctly predicted 61±7% of the same test molecules in a Turing-like test. Further, the computational routine provided insight into molecular features governing aggregation that had remained hidden to expert intuition. Leveraging our tool, we quantify that up to 15–20% of ligands in publicly available chemogenomic databases have the high potential to aggregate at typical screening concentrations, imposing caution in systems biology and drug design programs. Our approach provides a means to augment human intuition, mitigate attrition and a pathway to accelerate future molecular medicine.

## Introduction

Drug discovery is fuelled by small-molecules, either as tools to interrogate biology or as investigational leads towards therapeutics.[1-3] The successful development of such entities entails a transformative power on disease modulation, but a large proportion of them fail to reach the clinic due to attrition.[4] In that regard, it is increasingly recognized that a significant fraction of screening molecules present low aqueous solubility that manifests through nano/microscale agglomerates. These colloidal aggregates can bind unspecifically to proteins, inducing local denaturation and apparent inhibition.[5] While undirected interactions between small molecules and proteins are not desirable, most biological assays lack resolution to distinguish between 'pathological' and true target engagement; thereby incorrectly presenting the aggregating molecule as a positive result. Indeed, colloidal aggregates account for the largest source of false positive hits in high-throughput screens, surpassing other well-documented 'con artists', such as the pan-assay interference compounds.[6,7]

Vendor molecules, approved drugs and natural products are used to query biology on a systems level. However, given the widespread potential for small molecule promiscuity[8-10] and aggregation,[6,11-13] assay recommendations have been made to scrutinize screening hits and de-prioritize chemical matter displaying unspecific target modulation.[14] In particular, biochemical assays – with and without detergent – and dynamic light scattering (DLS) have become the gold standard technologies for interrogating aggregation.[14,15] While biochemical assays can be expensive and laborious, DLS is more accessible and allows determining the dynamic radius of particles in solution at a given concentration.[15] Still, it rapidly becomes cumbersome – both from an execution and data analyses perspectives – when scaled. As alternative, the pharmaceutical community continues to seek for heuristics expediting the detection of liable chemical matter[16-19] at the primary screen stage. 'Drug-likeness'[20] and the 'quantitative estimate of drug-likeness' (QED)[21] have been widely employed[22] to prioritize small molecules in early discovery programs, but were developed for disparate purposes. Albeit valuable, these composite metrics do not correlate with attrition; many approved drugs fail to comply with at least one of Lipinski's drug-like rules for oral absorption[23] and the QED overlooks the background distribution of molecular properties for non-drugs.[24,25] Therefore, tailored methods to swiftly and accurately flag nuisance small molecules are in high demand.[26] If reliable, those tools may provide practical research companions, limit laborious and expensive experimentation, can streamline the (de)-validation of screening hits and accelerate development pipelines by diminishing attrition at all stages.

The Aggregator Advisor[27] provides a solution to flag small colloidally aggregating molecules (SCAMs),[11] on the basis of chemical similarity and lipophilicity computations. Albeit viable, this tool presents limited applicability[11] on a wider scale due to an intrinsic inability to extrapolate chemical patterns. Similarity searches in the Aggregator Advisor provide insufficient generalizability, despite the available reference data – currently >12,500 structures.[11,27] In practice, many SCAMs will pass the filter unnoticed and, therefore, a confident *in silico* identification of nuisances in bioactive chemical space is technically unfeasible with the most commonly employed computational tool. Still, its source data has been utilized to build more sophisticated flagging systems, such as the SCAM detective[28] and ChemAGG,[29] which provide new vistas onto colloidal aggregation, although in disconnect to a paramount parameter – molecule concentration.

We here develop and validate a generally applicable machine-learning tool that more realistically gauges the impact of colloidal aggregation in small molecule discovery programs. We show that, by focusing on homogenous data in regards to screening medium identity, molecular concentration and colloidal aggregation annotations, we can boost predictive accuracy to 80% and consistently outperform available computational technologies and expert human intuition on an extensive out-of-sample validation set. Moreover, using state-of-the-art model interpretation, we identified both hitherto unknown chemical features correlated with molecular aggregation and pitfalls in the formalization of domain knowledge. Further, a large literature screen revealed >1.1 million potentially erroneous ligand associations that may skew pharmacology networks. Taken together, our predictive pipeline may contribute to advance future molecular design by alleviating attrition, and thus reliably assist expert intuition in decision-making.

## Results and discussion

**Machine learning for SCAM detection.** We built a fully-connected deep learning classifier from readily available DLS data.[15] We elected to train a model based on this dataset because of its homogeneity and causal link between aggregation and a specified, typical primary screen concentration (30 µM), which is neglected by previous tools. While the Aggregator Advisor dataset is more comprehensive, it would provide an unsuitable training set, in the present case. This is due to an indiscriminate collection of data from different sources – *e.g.* DLS and detergent-based biochemical assays – and, more importantly, molecule annotations at different concentrations,[27] which does not map onto our measured endpoint. Together, these are expected to affect model performance and provide information suited to a distinct research question, *i.e.* wherein screening concentration is not relevant. Pathological target modulation can be identified running paired biochemical assays (with and without detergent) and aggregates might result from assay-specific components. However, prior contributions[11,15,27] have shown an excellent correlation between confirmed unspecific binding and particle formation under the DLS screening conditions utilized in our study. We thus deemed the DLS readings appropriate for modelling and being a motivated proxy to estimate promiscuous binding.

Curation of the DLS data, *i.e.* elimination of molecules originally labelled as 'ambiguous', provided the required training set comprising 916 entities annotated with an 'aggregation'/'non-aggregation' readout (30:70 ratio, respectively) at a typical primary screen concentration (30 µM). We calculated substructural Morgan fingerprints (2048 bits, radius 3) and topological physicochemical properties ($n$ = 199; RDKit) for each molecule to abstract chemical connectivity and obtain a high dimensional descriptor vector. We then implemented an exhaustive parameter search routine to identify a classifier with optimal performance (*cf.* Methods and Supporting Figure S3–4).[30] As assessed through stratified 10-fold cross-validation studies, a feed-forward neural network with 3 hidden layers (DeepSCAMs, Figure 1a,b) was selected for downstream studies. In parallel, we ran adversarial controls[31] to de-validate the soundness of our classifier. By randomizing the target class annotations, and subsequently building an optimized neural network we observed that model performance decreased, and became comparable to random guessing from our training set (Figure 1b). This supports the disruption of relevant patterns during the Y-shuffling process, the meaningfulness of the employed abstractions, and the potential utility of our native model. In part, these results rule out overfitting, which is a realistic concern for neural networks and high dimensional descriptors with unusually small, yet real world datasets.

To prospectively challenge the utility of our machine-learning model, we queried >1.2 million purchasable screening molecules, and predicted their probabilities of generating colloidal aggregates. For experimental validation, candidate selection was performed by employing the MaxMin algorithm to ensure a chemotype-agnostic, diverse selection while covering the whole range of prediction probabilities as either 'aggregators' or 'non-aggregators' (*cf.* Supporting Table S1). In total, we procured 65 screening molecules and executed DLS screens to monitor both particle formation and auto-correlation curves at a concentration of 30 µM (Figure 1c). Based on the collected data, we confirmed 80% of correct predictions, which is fully aligned with the cross-validation studies. Not only did DeepSCAMs present a near perfect performance for high confidence predictions (19 in 20 correct predictions when class probability >95%) but also good performance for lower confidence outputs (33 in 45 correct predictions), supporting the ability to extrapolate learnt patterns. Importantly, the average Tanimoto index between the 65 screening molecules and their nearest neighbours in the DeepSCAMs training set is only 0.20±0.09 (Figure 1d). An arbitrary Tanimoto index cut-off can be set between 0.65 (*e.g.* ref. [32]) and 0.8 (*e.g.* ref. [27]). In all our prospective validation molecules, the similarity to the nearest neighbour falls under the cut-off, and no apparent relationship to the predicted class probability was found (*cf.* Supporting Figure S2). We also noted that the selected chemotypes extend beyond the typical 'drug-like' space in the training set [$QED_{test_molecules} = 0.52 \pm 0.18$ ($n$ = 65) *vs.* $QED_{training_set} = 0.63 \pm 0.18$ ($n$ = 916); $p = 2.28 \times 10^{-5}$, Welch *t*-test].[21] Together, these observations attest to the substructural dissimilarity between the validation and training sets, motivate the former as non-trivial use cases from a machine learning vantage point and thus, appropriate for challenging the generalizability / domain of applicability of DeepSCAMs.

We next contextualized the utility of DeepSCAMs by employing four optimized machine-learning algorithms on our training data (*k*NN, decision trees, random forests and AdaBoost; *cf.* Supporting Table S2). We found that none of those alternative methods presented better predictive performance (*cf.* Supporting Figure S1), pinpointing the superior ability of DeepSCAMs in extracting complex data patterns. Our method also outperformed the baseline flagging tool – Aggregator Advisor (*cf.* Supporting Table S1) – by correctly classifying a large proportion of the confirmed aggregators in the out-of-sample test set (recall of 76% for DeepSCAMs *vs.* 29% for Aggregator Advisor; Figure 1e; $p$ = 0.05 for Aggregator Advisor *vs.* $p = 1.17\times10^{-6}$ for DeepSCAMs, two-sided binomial test against random label selection). The result highlights an important caveat of similarity searches. They depend on the number and diversity of reference molecules, thus leading to a large fraction of false negatives if the chemical space is not broadly covered. Finally, our model also compared favourably relative to ChemAGG (Figure 1f and Supporting Table S1; $p = 2.62\times10^{-3}$, two-sided binomial test)[29] and the SCAM detective[28] (Supporting Table S1). We hypothesized that the enhanced generalizability of DeepSCAMs relative to other reported methods was linked to a leaner (<10% in size to either ChemAGG and SCAM detective) and more homogeneous training set. Indeed, we found no benefit in training DeepSCAMs with the Aggregator Advisor reference set. Because this data is highly imbalanced (95% aggregators) we downsampled the majority class to obtain an even distribution of examples – *i.e.* 653 molecules per class. Training led to a classifier with accuracy of 75% ($n$=3), which could be decreased by shuffling the target variables (50%, $n$=3). These control computations show that curating data with an experimental physical-organic chemistry motivation is an advantageous strategy that had previously been overlooked. The tailored training set provided the basis for augmenting machine perception,[33] and a preferential means of answering a research question that cannot be tackled otherwise.

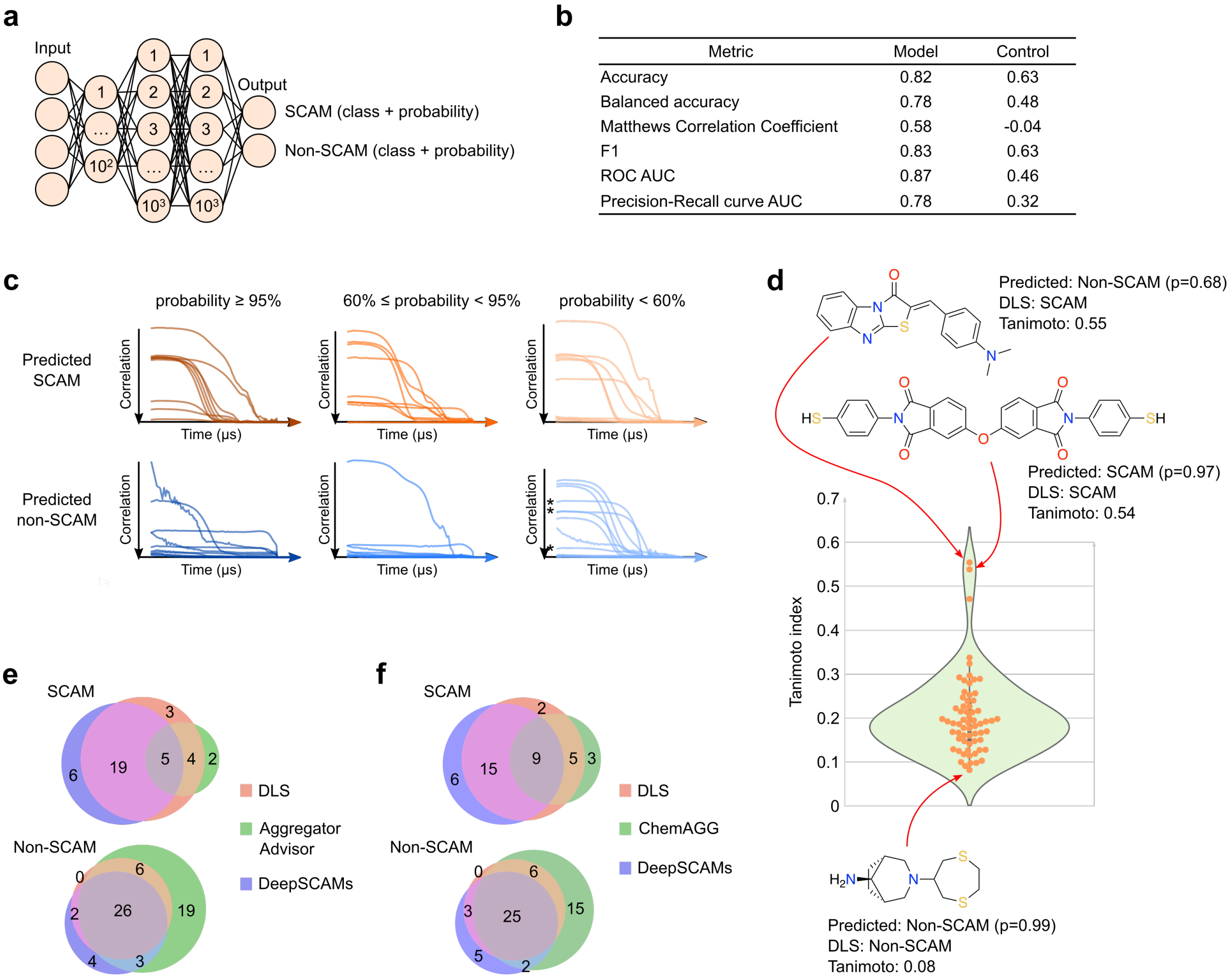

| Metric | Model | Control |
|---|---|---|
| Accuracy | 0.82 | 0.63 |
| Balanced accuracy | 0.78 | 0.48 |
| Matthews Correlation Coefficient | 0.58 | -0.04 |
| F1 | 0.83 | 0.63 |
| ROC AUC | 0.87 | 0.46 |
| Precision-Recall curve AUC | 0.78 | 0.32 |

**Fig. 1** Implementation and validation of a feed-forward deep neural network for flagging small colloidally aggregating molecules (SCAMs). a) Schematics of the optimal neural network architecture (DeepSCAMs). The input [2247 dimensional descriptor vector, including Morgan fingerprints (2048 bits, radius 3) concatenated with 199 RDKit topological descriptors] is fed into 3 hidden layers (100, 1000 and 1000 neurons, respectively). A class prediction ('SCAM'/'non-SCAM') and its probability are obtained as output. b) Model quality metrics as assessed through stratified 10-fold cross-validation studies for an imbalanced dataset containing 30% of positive cases (confirmed aggregators at 30 μM) and 70% of negative cases (confirmed non-aggregators at 30 μM). Data shows robust performance relative to control models obtained in Y-randomization tests. c) Experimental validation (auto-correlation curves) of 65 diverse small molecules through DLS screen. A sigmoidal behaviour denotes particle formation and/or sedimentation. Molecules are binned according to the predicted class and prediction confidence. *No particles detected. d) Distribution of structural similarity (Tanimoto index) between the 65 screened molecules and the nearest neighbour in the DeepSCAMs training set. Data shows that relevant patterns are identified independently of the similarity. Average Tanimoto index = 0.20±0.09. Examples of screened molecules are depicted, together with the predicted class and probability ('p'; see Supporting Figure S2 for correlation between Tanimoto and prediction probability values). e) Comparison of performance between DeepSCAMs (blue) and the Aggregator Advisor (green), using the 65 screened molecules for benchmarking. DLS (orange) corresponds to the ground truth. Data shows higher performance of DeepSCAMs for the detection of both SCAMs and non-SCAMs. f) Comparison of performance between DeepSCAMs (blue) and ChemAGG[29] (green), using the 65 screened molecules for benchmarking. DLS (orange) corresponds to the ground truth. Data shows higher performance of DeepSCAMs for the detection of both aggregators and non-aggregators.

**Chemical knowledge through model interpretation.** Despite the prospective utility and favourable performance of DeepSCAMs relative to other *in silico* methods, we wondered if our software tool could correctly formalize domain knowledge and augment perception on established, yet unwritten rules of chemical intuition. To that end, we enquired a panel of expert, PhD-level medicinal chemists and chemical biologists from academia and industry in Europe, Asia and the USA in a Turing-like test. The 15 experts were instructed to label each of the 65 atypical (*cf.* QED values) screening molecules as 'SCAM'/'non-SCAM', considering a test concentration of 30 µM in buffered aqueous solution (pH 7). The results clearly show that DeepSCAMs has a competitive vantage point relative to the chemists and more efficiently sieved through hidden chemistry patterns. Specifically, DeepSCAMs was not only able to correctly classify cases that were consensual for the expert panel but also others where the majority of the chemists failed (Figure 2a,b). Interestingly, the observed general disagreement between the surveyed experts with distinct clusters of domain knowledge reiterates how subjective and intuition-based the prediction of aggregation is (*cf.* Supporting Figure S9). Conversely, DeepSCAMs offers a robust, unbiased and efficient solution to flag SCAMs beyond intuitive binary rules.

To gauge the impact of each molecular feature for the DeepSCAMs output and shed light into the prediction disparities relative to expert intuition, we calculated Shapley values from game theory, as a solution maintaining both local accuracy and consistency. In short, Shapley values quantify the cooperative impact of a given feature as the change in the expected value to the model's output when a feature is observed versus the unknown[34] (Supporting Note S1). Deconvolution of our machine-learning model into such quantitative metrics showed the multidimensional nature of small molecule aggregation and motivated (dis)agreements with the expert chemists' predictions (Figure 2b,c). Specifically, it revealed that sulfonamides can promote solubility, and that flexible and lipophilic amines can provide the blueprints for aggregation – a realization that was either overlooked or unnoticed, as only one of the enquired chemists was able to correctly classify the molecule in question (Figure 2b). We challenged the relevance of this hitherto unknown trend through repeated SHAP analyses, and use of LIME[35] as a complementary approach (*cf.* Supporting Figure S5-8) to provide statistical support and reveal robustness in the model interpretation. Interestingly, both our tool and the majority of consulted experts failed to recognise patterns leading to either aggregation or non-aggregation in a couple of small molecules (Figure 2b). The result shows there are a number of relevant aggregation signatures that remain unknown; these can be unravelled further as data in contiguous search spaces becomes available. The same reasoning can be applied to the nonsensical predictions we obtained for more established test cases, where a significant proportion of experts were able to correctly assign a label. In some of those molecules, the DeepSCAMs prediction did not deviate significantly from random guessing (*i.e.* near misses), but still failed to recognise molecular shape and conformations that could promote one or the other class. Despite the marginal failures discussed herein, the high overall accuracy in out-of-sample predictions attest to the value of DeepSCAMs, wherein the model can establish data correlations that are currently unexpected/unapparent.

From a model-wide vantage point, we identified that the number of relevant features largely exceeds the typical number – four – efficiently processed by humans,[36] which may help explaining the competitive performance of DeepSCAMs. Indeed, our data corroborates that substructural fingerprints are appropriate descriptors to identify SCAMs, as some bits are highly positively or negatively associated with the probabilistic output (Figure 2b,c). Curiously, calculated log*P* – a property of paramount importance for the Aggregator Advisor – did not rank among the top-20 features in this instance. The result however does not de-validate the importance of log*P*, which is expected to play a more or less important role for explaining individual predictions. Rather, it demonstrates that flagging molecules based on a single calculated physicochemical descriptor may be an over-simplistic approach, leading to mispredictions. Conversely, molecular partial charges and the number of aliphatic carbocycles more significantly impact on model performance. For example, a high number of aliphatic carbocycles lowers the probability of a molecule being predicted as SCAM (Figure 2c), as they induce bulkier shapes and decrease stacking interactions.[37] For enhanced interpretability, we next projected the decision path of DeepSCAMs – via the corresponding activation outputs per layer – to reveal that SCAMs and non-SCAMs in training (916 molecules) and test data (65 molecules) become increasingly well separated along the neural network depth, despite originally overlapping in the input space (Figure 2d). This not only depicts the complexity of the problem in hand – wherein both classes are naturally intermixed – but also justifies the chosen model architecture and rationalizes the prospective utility of the tool. A close inspection of the projected data revealed opportunities for method development, as some molecules were poorly separated from the opposite class neighbours', irrespective of being correctly or misclassified (*cf.* Supporting Figure S3–4 and Table S3–4). Overall, our approach provides an intuitive and readily visualizable reasoning on the DeepSCAMs ability to formalize chemical knowledge (*cf.* selected molecules in Figure 2). Because we were able to delve the inner workings of DeepSCAMs, to build trust on its architecture and obtain interpretable/explainable decision processes that are largely aligned with expert intuition, we next expanded the application of DeepSCAMs to bioactive compound databases in hope of illuminating liable matter.

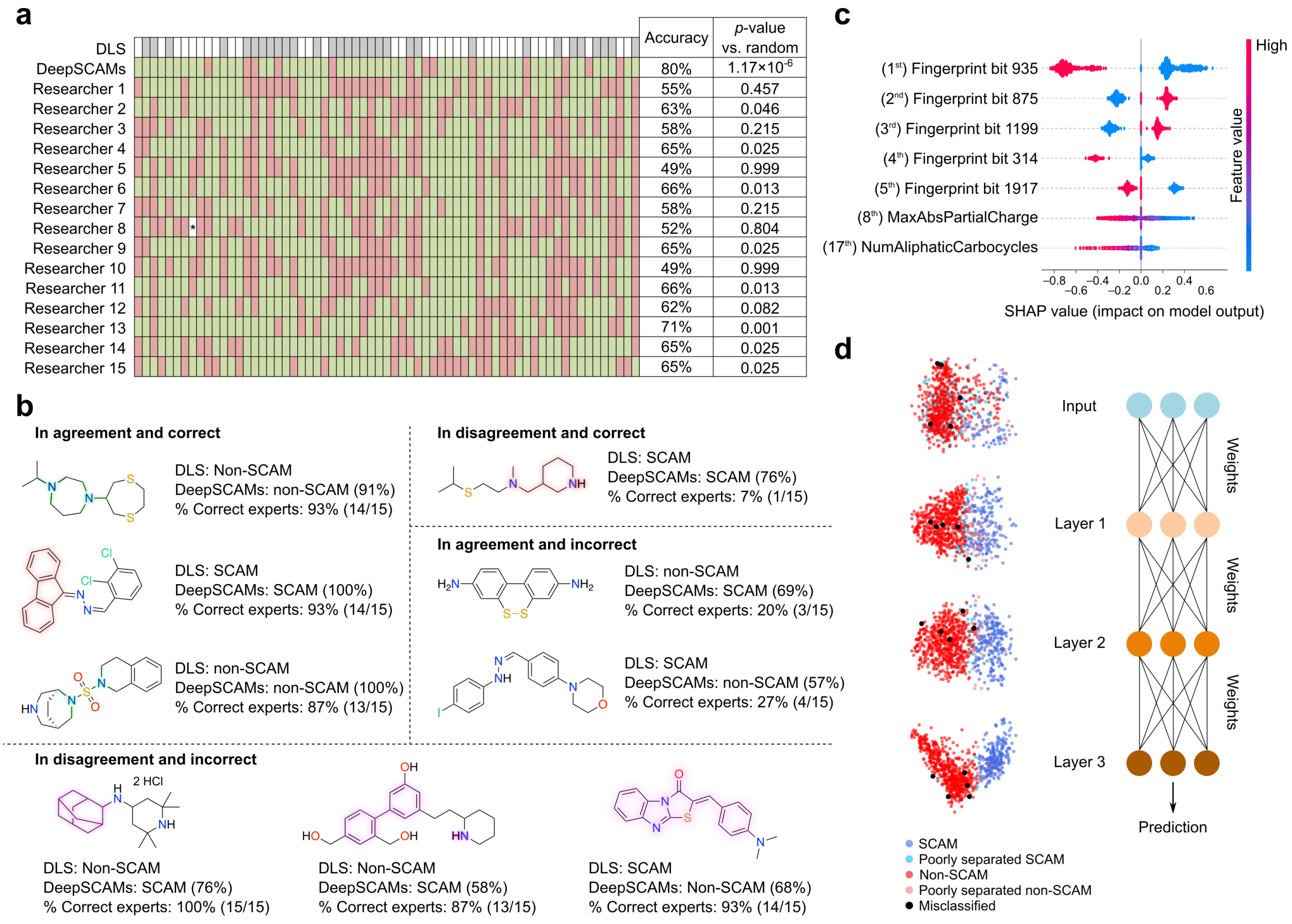

**Fig. 2** Formalization of expert intuition. a) Prediction comparison between DeepSCAMs and 15 expert chemists from academia and industry. The top row denotes the experimental result (ground truth) from the DLS screen at 30 µM (pH = 7), followed by DeepSCAMs and the researchers' answers; Grey: SCAM, White: non-SCAM, Red: incorrect prediction, Green: correct prediction, Asterisk: missing answer. The percentage of correct predictions is provided as well as a statistical analysis versus random guessing (two-sided binomial test). b) Structure of selected molecules that were correctly or incorrectly predicted as SCAM or non-SCAM by both DeepSCAMs and expert chemists. In multiple cases, the algorithm and chemists are in agreement and in other cases in disagreement. For predictions correctly made by both software and chemists the structural features are highlighted in red (supporting SCAM) and green (supporting non-SCAM). For a correct prediction by DeepSCAMs in disagreement with the expert panel, the relevant structural features are highlighted in orange. Collectively, those features are underrepresented in the training set (2 molecules), which may explain the responses given by the chemists. For molecules mispredicted by DeepSCAMs, but correctly labelled by the experts, the highlighted motifs represent consensual features as per human intuition ($n$=5). c) Model interpretation (based on training data) showing the top-5 global features using Shapley values. The top-2 physicochemical properties are also provided as well as their global ranks in parenthesis. Data shows that different fingerprint bits have varying weights on the model output (prediction as SCAM) and that both a high number of aliphatic carbocycles and molecular partial charges are negatively correlated with prediction as SCAM. d) Isomap of activation outputs for each of the three layers in DeepSCAMs. The projection shows that training data becomes better separated with the neural net depth. Prospective examples that were incorrectly labelled by DeepSCAMs, but not by the majority of surveyed chemists are highlighted (black dots). For details on how to distinguish well- from poorly-separated data please see Supporting Figure S3–4 and Table S2–3.

**Identification of biases in the literature.** Previously, *ca.* 7% of all compounds in the medicinal chemistry literature had been flagged by the Aggregator Advisor.[27] Albeit experimentally unconfirmed, such a large fraction of nuisance compounds can perturb exploratory and development pipelines. Flagging molecules as early as possible is desirable and can alleviate the pursuit of unfruitful research lines, but the lack of concentration–effect interdependence in the Aggregator Advisor output – as discussed above – makes gauging the impact of aggregation in primary screens currently inaccessible. Considering the higher generalizability of DeepSCAMs, we re-interrogated the prevalence of potential aggregators in >1.8 million ChEMBL[38] molecules, with the goal of providing machine-learnt insights onto SCAMs across the chemical biology literature, when and if employed in assays at 30 µM. As previously discussed, this is a realistic scenario in primary screens with the goal of identifying hits for medicinal chemistry elaboration. While accounting for the predictive uncertainty in DeepSCAMs, we found that up to 15–20% of all chemical entities are anticipated to aggregate at 30 µM (Figure 3) and can mislead biological investigations. The result is 2- to 3-fold higher than a prior estimate,[27] but is aligned with data from a small-sized random screen.[15] On a more extreme case, 95% of screening hits have shown pathological target engagement,[6] including not only colloidal aggregation, but also auto-fluorescence and reactivity with assay components. Our computations thus confidently support a higher prevalence of SCAMs than previously estimated, and provide a data-motivated interpretation for the qualitatively known,[39] but often neglected bias for harnessing nuisance compounds in molecular medicine. Apparently, this is an established and deep-rooted trend, as computed by DeepSCAMs in a time series analysis of ChEMBL 2009–2019 (Figure 3). The predicted SCAMs have a high degree of polyaromatic moieties (97% of all molecules), which endorse an impaired kinetic solubility. One possible explanation for this observation is the ease of synthesis of such motifs, relative to *sp*$^3$ hybridization-rich frameworks, to which the recent surge of C–C bond formation reactions – such as Suzuki-Miyaura coupling-type chemistry[40] – contribute to. The trend was corroborated by analysis of the BindingDB,[41] wherein 20% of the database could be predicted to aggregate at 30 µM while using our learning algorithm.

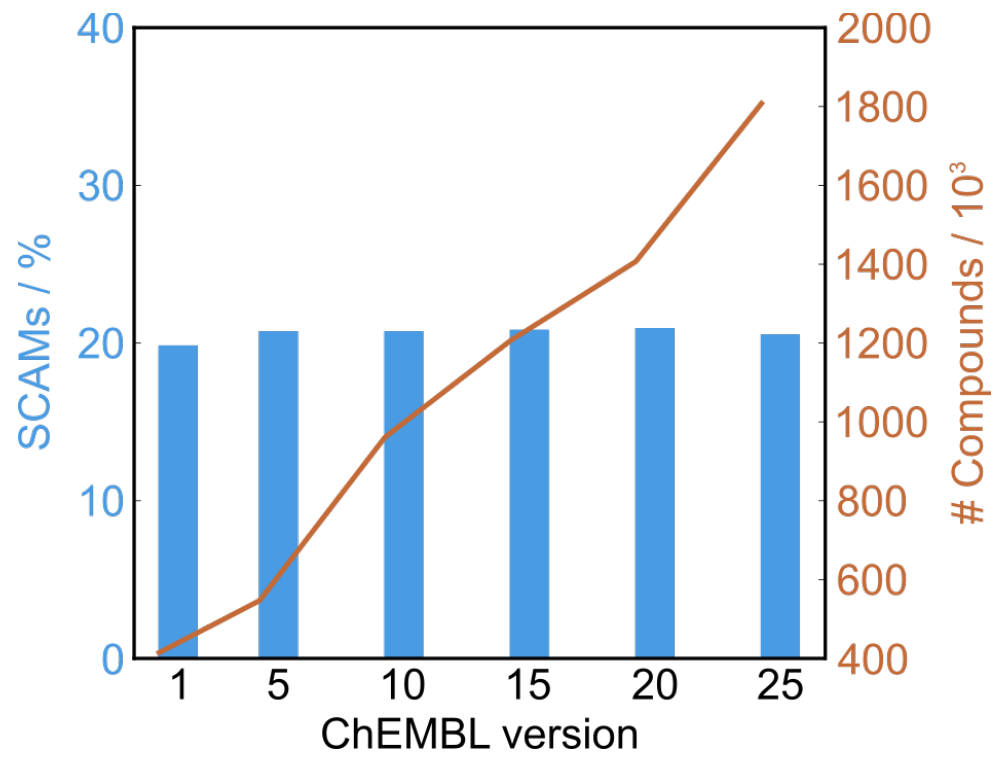

**Fig. 3** Predicted colloidal aggregation across different versions of ChEMBL. The results suggest deep-rooted biases in the exploration of potentially attrition prone chemical matter (20–21% of aggregators across the time series).

Given the use of databases, such as ChEMBL and BindingDB, in model implementation and structure-activity-relationship analyses, this result is alarming for the medicinal chemistry and chemical biology communities, as SCAMs might misguide model development or mislead hit and lead discovery programs. We next extracted protein and activity annotations for each entity, as reported in ChEMBL, to study the impact of predicted SCAMs at an integrated biology level. Our analyses suggest that all protein families are likely affected by false positive hits and that 5,733 predicted SCAMs have bioactivity annotations – against 531 distinct targets – above 30 µM. Our analyses also showed multiple high IC/$EC_{50}$ or $K_{D/i}$ values for molecules screened against isoforms of the cytochrome P450 (Figure 4a). Strikingly, the microtubule-associated protein tau and beta-secretase 1 are common targets of potential SCAMs. Both macromolecules have been explored in Alzheimer's disease[42-44] and the inclusion of entities with a questionable medicinal chemistry signature as prototypes for development might partly explain a number of unsuccessful early discovery programs in this space, together with convoluted biology. On a broader perspective, a total of 6,078 ligand–target relationships are likely compromised, which may incorrectly link 659 distinct targets (Figure 4b). Furthermore, >1.1 million predicted SCAM relationships endorsed by drug target commonalities were identified and, thereby, can inflate their respective ligandable spaces. Undoubtedly, "false hits" and assay interference are established notions,[4] but quantifying the impact of aggregation on discovery chemistry has been elusive, to the best of our knowledge. Despite the accuracy of DeepSCAMs, the data insights would require large-scale experimental confirmation. Still, our machine-learning tool frames the colloidal aggregation problematic in an unprecedented way, which has profound implications *per se*. Overall, it reinforces an urgent need for enforcing standardized data quality measures. The implemented measures must be widely adopted, including routine biochemical counter-screens in presence and/or absence of detergents or DLS experiments. Notoriously, similar recommendations have been made,[39] but a relatively narrow adoption seems insufficient to revert the observed trends. Our quantitative data resonates with those calls and motivates the development of higher quality probes and drug leads, as the means to gradually reduce the impact of SCAMs in discovery sciences, and more efficiently advance exploratory and late-stage development pipelines.

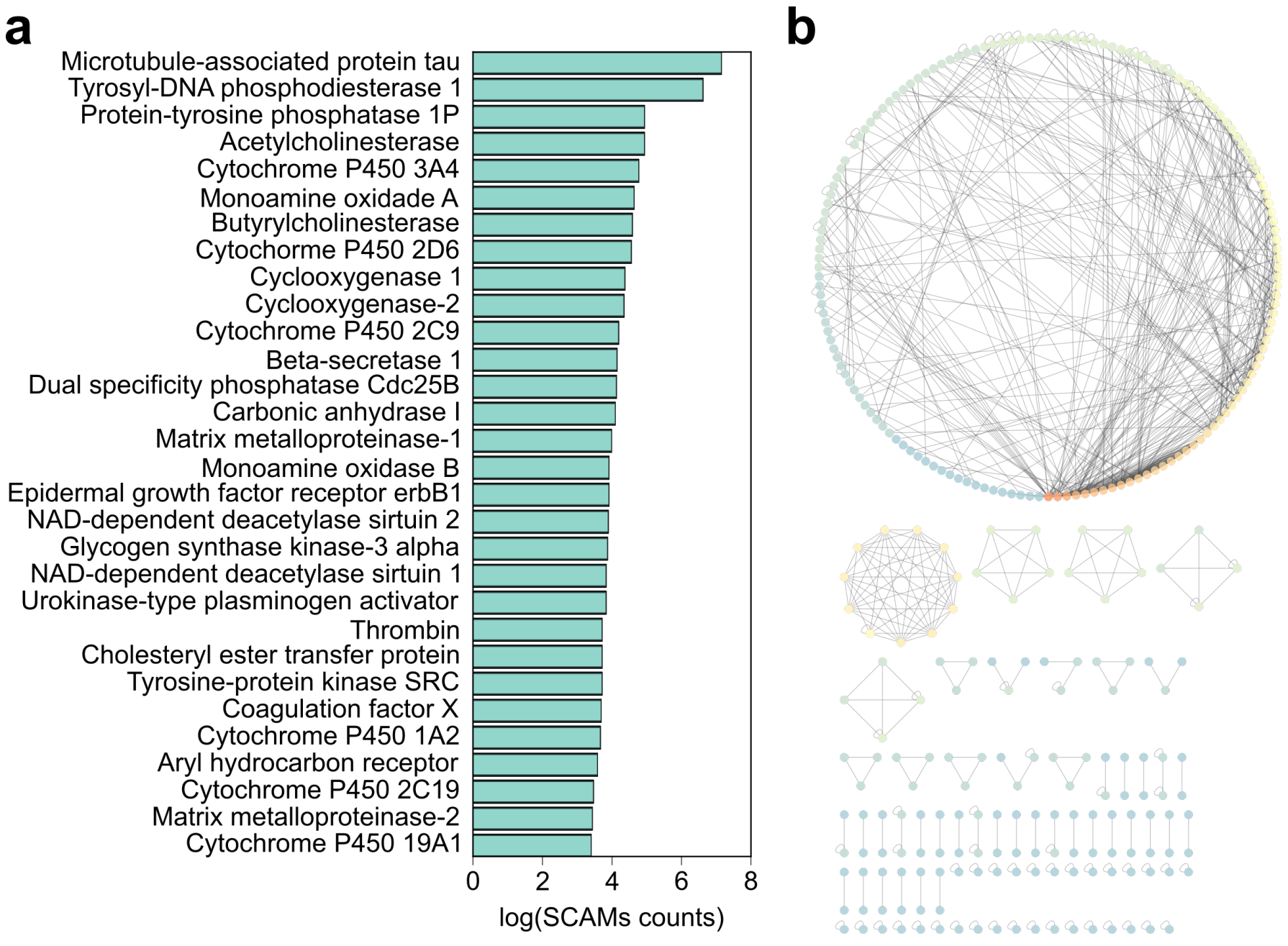

**Fig. 4** SCAMs interfere with network pharmacology. a) Predicted SCAMs ($log_{10}$Counts) in ChEMBL display $IC_{50}$, $K_D$ or $K_i$ values >30 μM for a wide range of proteins. The top-30 targets more likely affected by false positive readouts elicited by SCAMs are displayed; these include proteases, enzymes, kinases. b) False positive readouts generate skewed pharmacological networks that might be currently explored in early discovery programs. The network describes target correlations (nodes) via SCAMs commonalities (edges; $IC_{50}$, $K_D$ or $K_i$ value >30 μM). Color codes depict the connectivity number; blue = low connectivity, orange = high connectivity. The network was built with Cytoscape 3.4.0.

## Conclusions

In summary, we built and prospectively validated a deep neural network to flag SCAMs. While concentration and screening medium identity are key in governing the colloidal aggregation of small molecules, all available *in silico* technologies neglect their influence towards the binary output. Conversely, DeepSCAMs leverages homogeneous data and indirectly formalizes, for the first time, said physical-chemical correlation to reveal an unprecedented utility of machine learning. Our tool outperformed preceding computational workflows and human experts on a challenging prospective test. Yet, it might be improved further as more standardized DLS data is collected. Importantly, we harnessed machine learning to generate new chemical knowledge. The statistical model confidently unveiled chemical features – such as lipophilic amines – that influence small molecule aggregation and have so far remained overlooked by most expert medicinal chemists. Moreover, DeepSCAMs exposed persistent and alarming biases in the literature that favour liable chemical matter. If experimentally uncontrolled for, those SCAMs can misguide rational drug design across all target families, with higher-than-envisaged consequences in present and future small molecule development. With the advent of automated, data-driven research workflows,[45,46] and the overall importance of small molecules in drug discovery, we expect the propagation of misleading structure–activity correlations to further disrupt development pipelines, endorsing unfruitful drug investigations and slowing down the identification of life-changing therapies. DeepSCAMs and related tools can play a crucial role in augmenting domain intuition, probabilistically informing decision-making and safe-guarding the design of bioactive matter, thereby leveraging pre-clinical efficiency and supporting next generation molecular medicine.

**Acknowledgements.** G.J.L.B. is a Royal Society URF (URF\R\180019). T.R. is an *Investigador Auxiliar* supported by FCT Portugal (CEECIND/00684/2018). The authors thank V. Lunot for valuable discussions.

**Competing interests:** K.L. and A.Y. are employees of Insilico Medicine Taiwan. Y.-C.L. is CEO of Insilico Medicine Taiwan. G.J.L.B., and T.R. are co-founders and shareholders of TargTex S.A.

**Author contributions:** K.L., A.Y. and D.R. contributed with machine learning models and data analyses. Y.-C.L. and G.J.L.B. analysed data and supervised research. T.R. implemented DeepSCAMs, performed screens and analysed data. T.R. conceived, designed and coordinated the study. T.R. wrote the manuscript with contributions from the remaining authors. K.L. and A.Y contributed equally to this study. All authors agreed on submitting the current version of the manuscript.

# Supporting Information

Combating small molecule aggregation with machine learning

## Materials and Methods

***Machine learning and data analyses***. All software was implemented in Python 3.7, using the NumPy 1.11.3, Pandas 0.19.2, RDKit 2020.03.1 and Scikit-learn 0.18.1 libraries. DeepSCAMs was constructed with publicly available DLS data,[1] using a full grid hyperparameter search [alpha = 0.0001/0.001; iterations = 200/500/1000; activation=logistic/tanh; solver = lbfgs/sgd; learning rate = constant/adaptive; hidden layers = 1–3 layers with 10/100/1000 neurons] and stratified 10-fold cross-validation for model selection. The selected feed-forward neural network uses 3 hidden layers with 100, 1000 and 1000 neurons, respectively, alpha value of 0.0001, tanh as activation function, stochastic gradient descent and constant learning rate over 200 iterations. All remaining hyperparameters were used as default. Data analysis pipelines were performed with SciPy 1.2.3, plotted in Matplotlib 1.5.3, Matplotlib-Venn 0.11.5 or Seaborn 0.10.1. Figures were compiled in Inkscape 0.91. Computations were carried out on an Apple Mac Pro machine (3.5 GHz processor, 32 Gb RAM) or Google Colab. Code is available at https://github.com/tcorodrigues/DeepSCAMs. The Chembridge CORE/ExpressPick, ChEMBL v1–25[2] and BindingDB[3] libraries were queried with DeepSCAMs. Note: while 'aggregation' measured through DLS can be used to flag promiscuous behaviour – and has shown good correlation with biochemical assay outputs[4] – it does not ensure pathological target engagement *per se*. Specifically, DeepSCAMs focuses on predicting DLS outputs – *i.e.* aggregation in a physiologically-relevant aqueous medium – and does not account for aggregation induced by assay conditions, such as presence of specific biomacromolecules.

***Dataset preparation.*** Publicly available DLS data[1] at a concentration of 30 μM was used for training. Curation consisted of checking the corresponding SMILES string validity (with RDKit) and discarding all molecules with 'ambiguous' as annotation. This resulted in a training set with 916 entities with an 'aggregation'/'non-aggregation' label and class imbalance ratio of 30:70, respectively.

***Candidate selection.*** Candidate selection for prospective validation was performed in KNIME using the RDKit MaxMin node. All molecules were empirically binned into three classes, according to the prediction probability (p): $p<60\%$, $60\leq p<95\%$, $p\geq 95\%$. 25, 20 and 20 molecules were selected from each group, respectively.

***Model interpretation.*** We employed SHAP 0.35.0[5] and LIME 0.1.1.37[6] to explain the DeepSCAMs predictions by training an interpretation model locally, as described.[5] The interpretation instantiates as a set of sparse important weights to indicate focal points of the model for their decision. For SHAP, we employed the kernel explainer (shap.KernelExplainer) and adopted *k*-means for dimensionality reduction and logit link. All remaining settings were used as default. For LIME, we employed tabular explanation models (lime.lime_tabular.LimeTabularExplainer) and set all 2048 fingerprint bits as categorical variables. All remaining settings were used as default. We repeated the analyses 20 times with different random state values to provide variance estimation for these linear explanatory models.

***Dynamic Light Scattering***. Stock solutions (30 mM) for each of the 65 purchased screening molecules were prepared in neat DMSO and subsequently diluted in filtered $KH_2PO_4$ (50 mM) to a final concentration of 30 μM. Data (three technical replicates) were collected on a Zetasizer Nano S (Malvern) at 25 °C. Classification as ‘aggregator’ required the visualization of an autocorrelation curve described by a sigmoid function and quantifiable particle sizes. These experimental conditions were reproduced from ref. 1, and had been utilized to screen the compounds in the training set. The ‘aggregator’/‘non-aggregator’ labels were used as target in the DeepSCAMs classifier.

***Survey.*** Fifteen medicinal chemistry experts from academia and the pharmaceutical industry in Europe, Asia and the USA were asked to assign an ‘aggregator’ or ‘non-aggregator’ label for each of the 65 assayed molecules, without knowledge of the DeepSCAMs and DLS output. Responses were voluntary and the anonymity of respondents was ensured. The responses were analysed by quantifying the number of mismatches relative to the ground truth (DLS data).

**Table S1.** Structures, predictions and colloidal aggregation (DLS) for the 65 test molecules.

| Code | Structure | DeepSCAMs (Probability) | SCAM Advisor (Tanimoto[a]) | ChemAGG[7] | SCAM detective[8] | Experimental DLS |
|---|---|---|---|---|---|---|
| A1 | 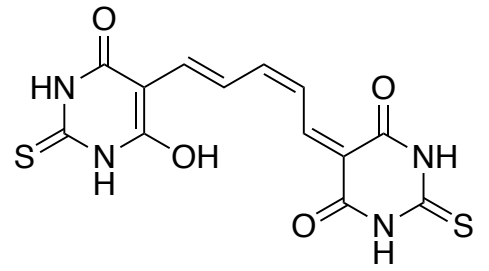 | **Non-SCAM** (0.97) | **Non-SCAM** | **Non-SCAM** | **Non-SCAM*** (0.80) | Non-SCAM |
| A2 | 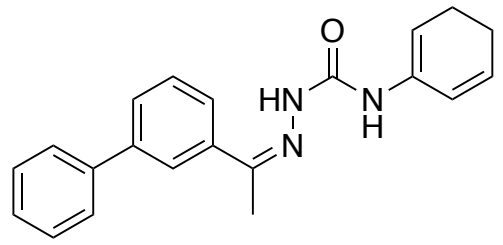 | **SCAM** (0.503) | Non-SCAM | **SCAM** | Non-SCAM* (0.50) | SCAM |
| A3 | 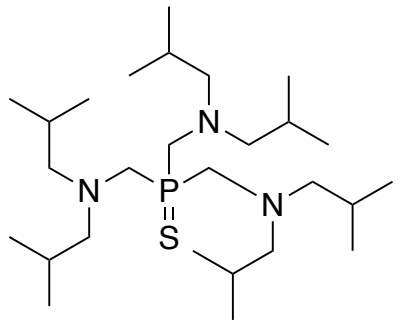 | **SCAM** (0.54) | Non-SCAM | Non-SCAM | Non-SCAM* (0.70) | SCAM |
| A4 | 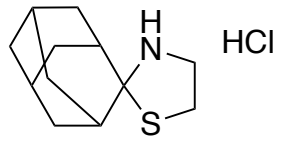 | **Non-SCAM** (0.54) | **Non-SCAM** | **Non-SCAM** | **Non-SCAM*** (0.70) | Non-SCAM |
| A5 | 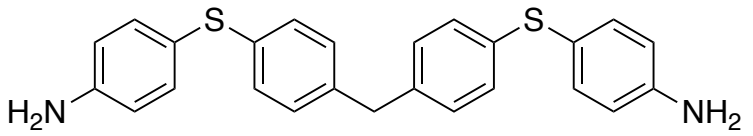 | **SCAM** (0.99) | Non-SCAM | Non-SCAM | Non-SCAM* (0.70) | SCAM |
| A6 | 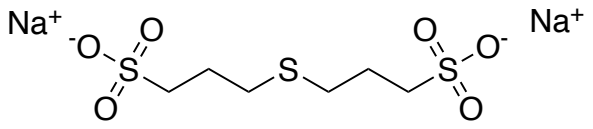 | **Non-SCAM** (0.99) | **Non-SCAM** | **Non-SCAM** | **Non-SCAM*** (0.90) | Non-SCAM |
| A7 | 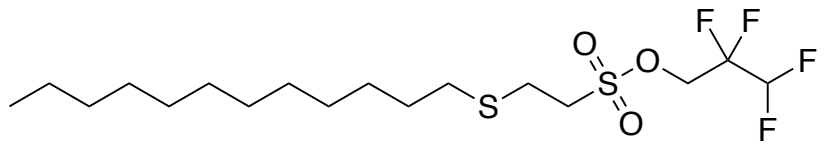 | **Non-SCAM** (0.60) | **Non-SCAM** | **Non-SCAM** | **Non-SCAM*** (0.60) | Non-SCAM |
| A8 | 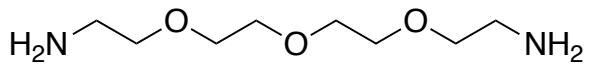 | **Non-SCAM** (0.98) | **Non-SCAM** | **Non-SCAM** | **Non-SCAM*** (0.80) | Non-SCAM |
| A9 | 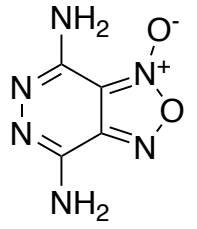 | **Non-SCAM** (0.999) | **Non-SCAM** | **Non-SCAM** | **Non-SCAM*** (0.80) | Non-SCAM |
| B1 | 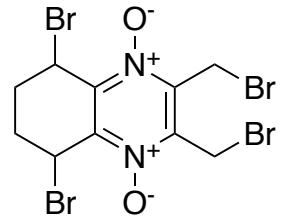 | SCAM (0.56) | **Non-SCAM** | **Non-SCAM** | **Non-SCAM* (0.80)** | Non-SCAM |
| B2 | 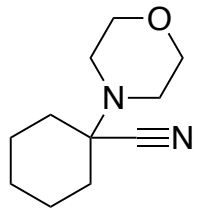 | **Non-SCAM (0.999)** | **Non-SCAM** | **Non-SCAM** | **Non-SCAM* (0.60)** | Non-SCAM |
| B3 | 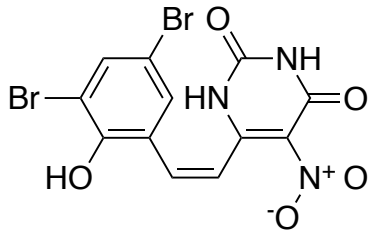 | **SCAM (0.53)** | Non-SCAM | Non-SCAM | Non-SCAM* (0.90) | SCAM |
| B4 | 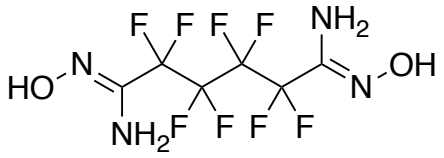 | **Non-SCAM (0.99)** | **Non-SCAM** | **Non-SCAM** | **Non-SCAM* (0.90)** | Non-SCAM |
| B5 | 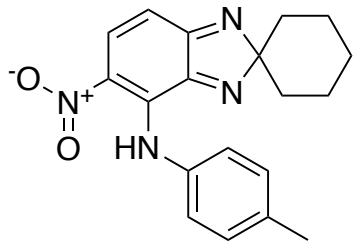 | **Non-SCAM (0.52)** | **Non-SCAM** | SCAM | **Non-SCAM* (0.70)** | Non-SCAM |
| B6 | 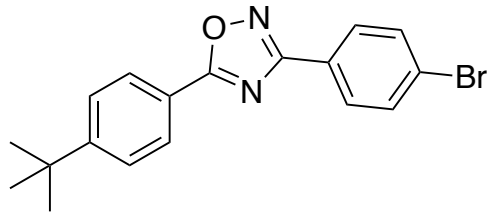 | Non-SCAM (0.56) | **SCAM (0.82)** | **SCAM** | Non-SCAM* (0.60) | SCAM |
| B7 | 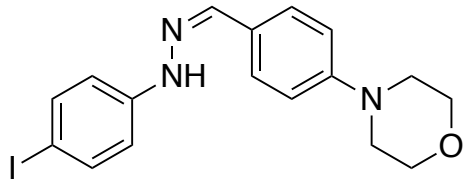 | Non-SCAM (0.57) | **SCAM (0.78)** | **SCAM** | Non-SCAM* (0.50) | SCAM |
| B8 | 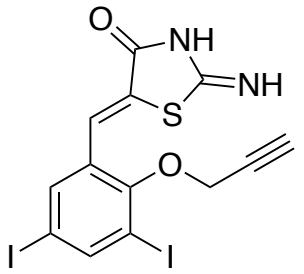 | **SCAM (0.997)** | Non-SCAM | Non-SCAM | Non-SCAM* (0.70) | SCAM |
| B9 | 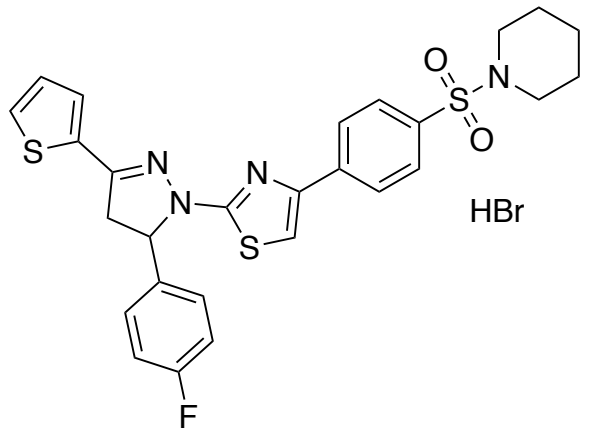 | **SCAM (0.97)** | Non-SCAM | **SCAM** | Non-SCAM* (0.50) | SCAM |
| C1 | 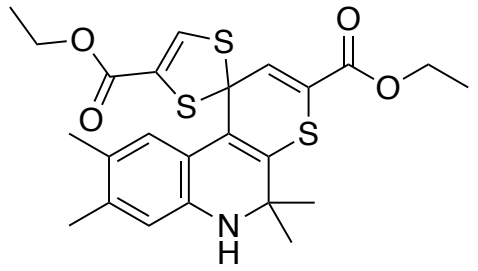 | Non-SCAM (0.51) | Non-SCAM | Non-SCAM | **SCAM*** (0.50) | SCAM |
| C2 | 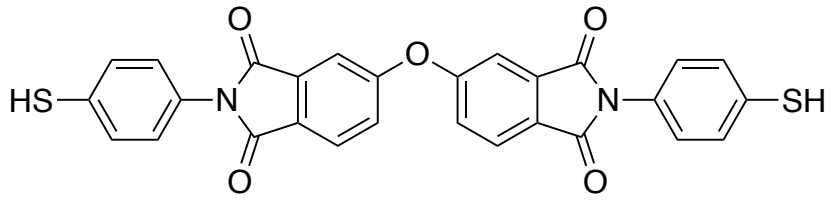 | **SCAM** (0.97) | **SCAM** (0.86) | Non-SCAM | Non-SCAM* (0.60) | SCAM |
| C3 | 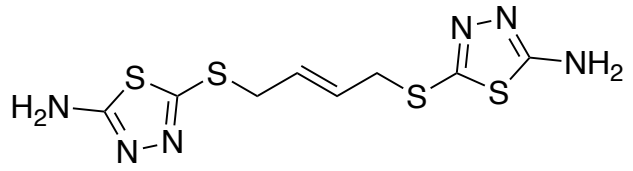 | **SCAM** (0.55) | **SCAM** (0.72) | Non-SCAM | Non-SCAM* (0.90) | SCAM |
| C4 | 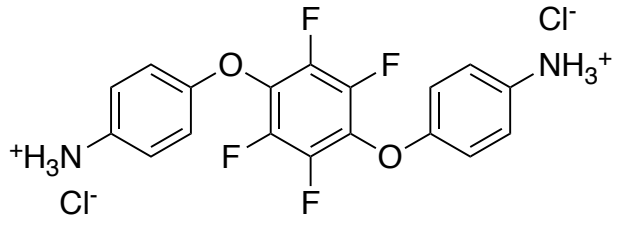 | **Non-SCAM** (0.55) | **Non-SCAM** | **Non-SCAM** | **Non-SCAM*** (0.80) | Non-SCAM |
| C5 | 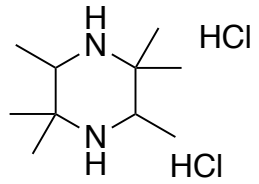 | **Non-SCAM** (0.97) | **Non-SCAM** | **Non-SCAM** | **Non-SCAM*** (0.80) | Non-SCAM |
| C6 | 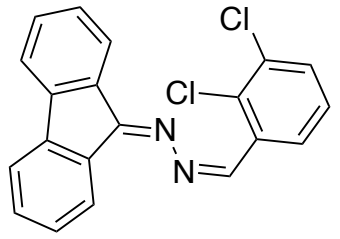 | **SCAM** (0.998) | Non-SCAM | **SCAM** | Non-SCAM* (0.70) | SCAM |
| C7 | 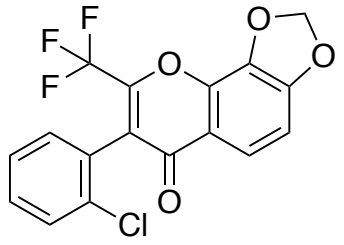 | **Non-SCAM** (0.55) | **Non-SCAM** | **Non-SCAM** | **Non-SCAM*** (0.70) | Non-SCAM |
| C8 | 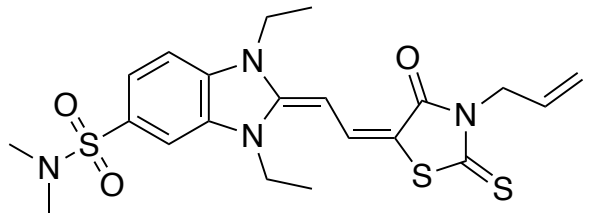 | Non-SCAM (0.57) | Non-SCAM | **SCAM** | Non-SCAM* (0.60) | SCAM |
| C9 | 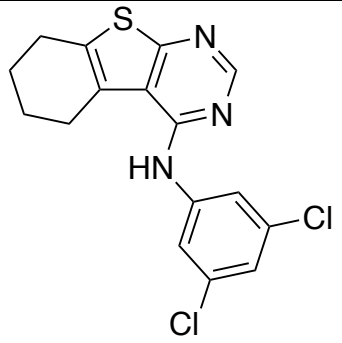 | **SCAM** (0.98) | **SCAM** (0.77) | **SCAM** | **SCAM** (0.50) | SCAM |
| D1 | 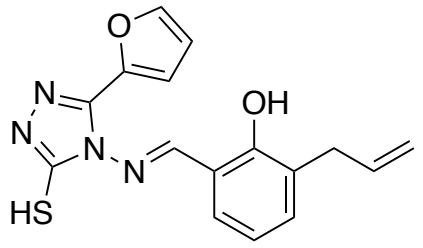 | **SCAM** (0.57) | Non-SCAM | **SCAM** | Non-SCAM* (0.60) | SCAM |
| D2 | 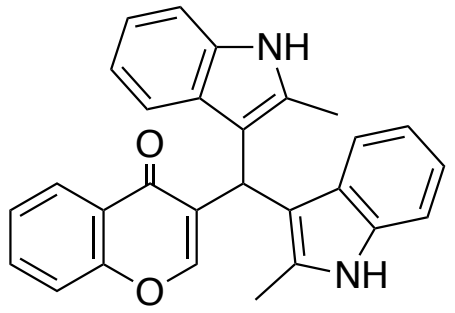 | **SCAM** (0.97) | Non-SCAM | Non-SCAM | Non-SCAM* (0.70) | SCAM |
| D3 | 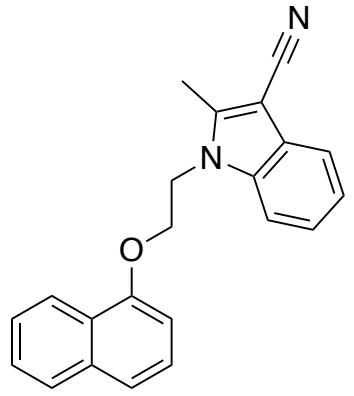 | **SCAM** (0.52) | Non-SCAM | **SCAM** | **SCAM** (0.50) | SCAM |
| D4 | 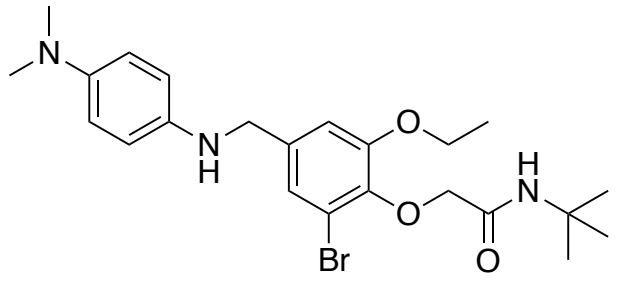 | **SCAM** (0.95) | **SCAM** (0.91) | **SCAM** | **SCAM** (0.80) | SCAM |
| D5 | 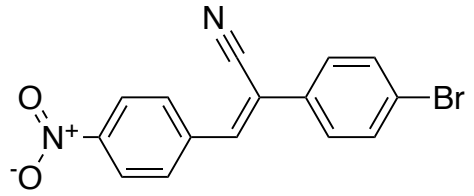 | **SCAM** (0.96) | Non-SCAM | **SCAM** | Non-SCAM* (0.80) | SCAM |
| D6 | 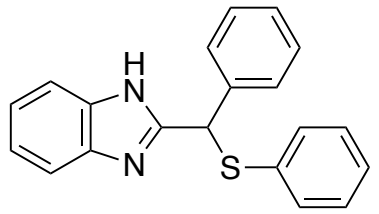 | **SCAM** (0.59) | Non-SCAM | Non-SCAM | Non-SCAM* (0.60) | SCAM |
| D7 | 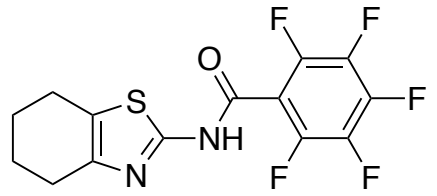 | **Non-SCAM** (0.51) | SCAM (0.80) | SCAM | **Non-SCAM** (0.90) | Non-SCAM |
| D8 | 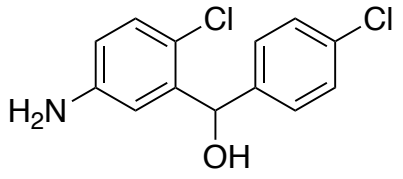 | **Non-SCAM** (0.53) | **Non-SCAM** | **Non-SCAM** | **Non-SCAM*** (0.70) | Non-SCAM |
| E1 | 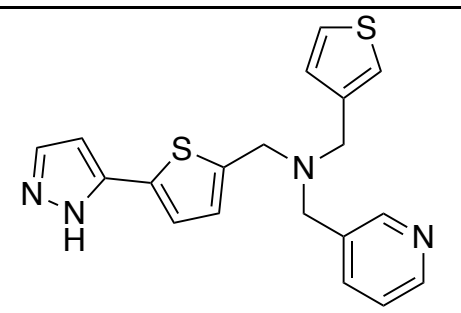 | Non-SCAM<br>(0.51) | Non-SCAM | Non-SCAM | Non-SCAM*<br>(0.60) | SCAM |
| E2 | 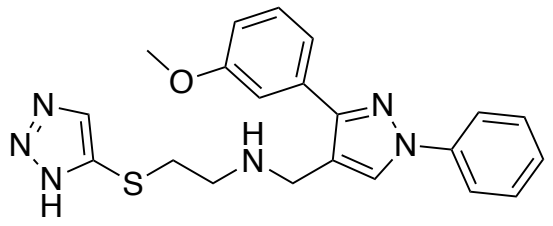 | **SCAM**<br>(0.99) | Non-SCAM | **SCAM** | **SCAM**<br>(0.70) | SCAM |
| E3 | 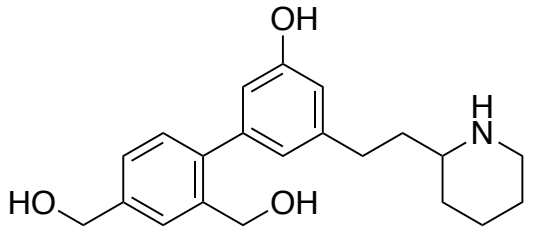 | SCAM<br>(0.58) | **Non-SCAM** | **Non-SCAM** | **Non-SCAM***<br>(0.50) | Non-SCAM |
| E4 | 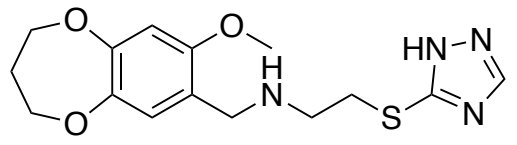 | SCAM<br>(0.51) | **Non-SCAM** | **Non-SCAM** | **Non-SCAM***<br>(0.60) | Non-SCAM |
| E5 | 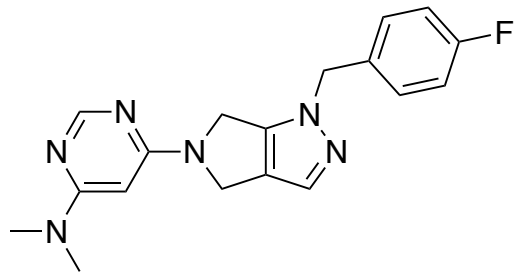 | **Non-SCAM**<br>(0.53) | **Non-SCAM** | **Non-SCAM** | SCAM*<br>(0.60) | Non-SCAM |
| E6 | 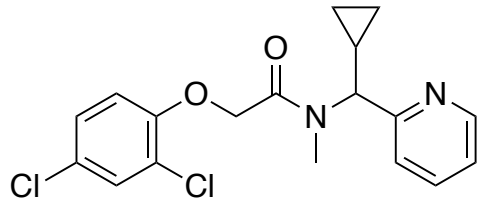 | **Non-SCAM**<br>(0.98) | **Non-SCAM** | **Non-SCAM** | SCAM<br>(0.60) | Non-SCAM |
| E7 | 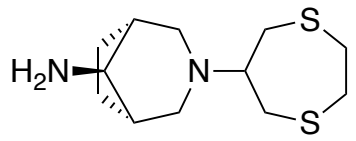 | **Non-SCAM**<br>(0.99) | **Non-SCAM** | **Non-SCAM** | **Non-SCAM***<br>(0.70) | Non-SCAM |
| E8 | 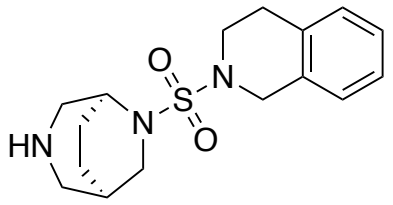 | **Non-SCAM**<br>(0.59) | **Non-SCAM** | **Non-SCAM** | **Non-SCAM***<br>(0.60) | Non-SCAM |
| E9 | 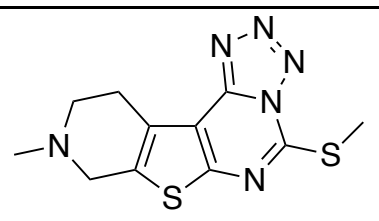 | Non-SCAM<br>(0.99) | **SCAM**<br>(0.73) | **SCAM** | Non-SCAM<br>(0.60) | SCAM |
| D9 | 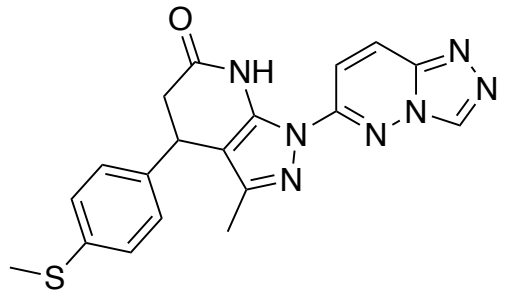 | **Non-SCAM** (0.59) | **Non-SCAM** | SCAM | **Non-SCAM*** (0.50) | Non-SCAM |
| B2_A1 | 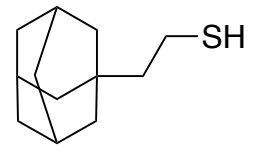 | **Non-SCAM** (0.94) | **Non-SCAM** | **Non-SCAM** | **Non-SCAM*** (0.60) | Non-SCAM |
| B2_A2 | 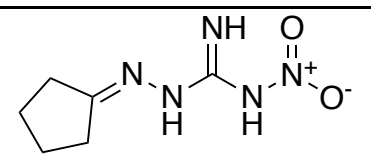 | **Non-SCAM** (0.94) | **Non-SCAM** | **Non-SCAM** | **Non-SCAM*** (0.80) | Non-SCAM |
| B2_A3 | 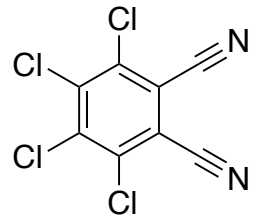 | **Non-SCAM** (0.84) | **Non-SCAM** | **Non-SCAM** | **Non-SCAM*** (0.90) | Non-SCAM |
| B2_A4 | 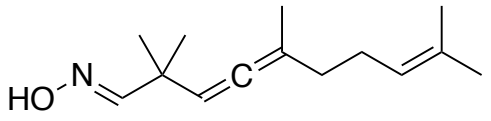 | **Non-SCAM** (0.88) | **Non-SCAM** | **Non-SCAM** | **Non-SCAM*** (0.70) | Non-SCAM |
| B2_A5 | 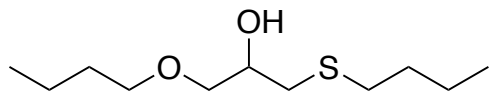 | **Non-SCAM** (0.89) | **Non-SCAM** | **Non-SCAM** | **Non-SCAM*** (0.60) | Non-SCAM |
| B2_A6 | 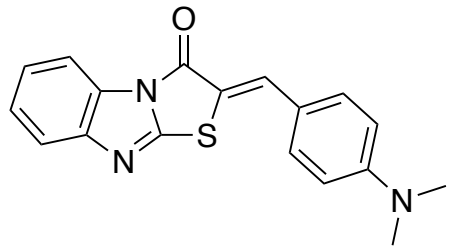 | Non-SCAM (0.68) | **SCAM** (0.80) | **SCAM** | Non-SCAM* (0.60) | SCAM |
| B2_A7 | 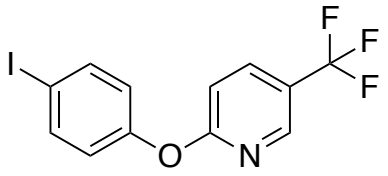 | **Non-SCAM** (0.83) | SCAM (0.76) | **Non-SCAM** | **Non-SCAM*** (0.80) | Non-SCAM |
| B2_A8 | 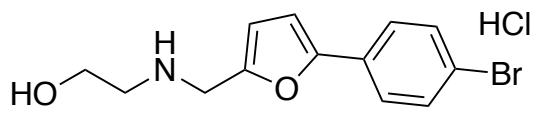 | **Non-SCAM** (0.93) | **Non-SCAM** | **Non-SCAM** | **Non-SCAM*** (0.50) | Non-SCAM |
| B2_A9 | 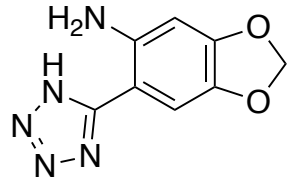 | **Non-SCAM** (0.90) | **Non-SCAM** | **Non-SCAM** | **Non-SCAM*** (0.80) | Non-SCAM |
| B2_A10 | 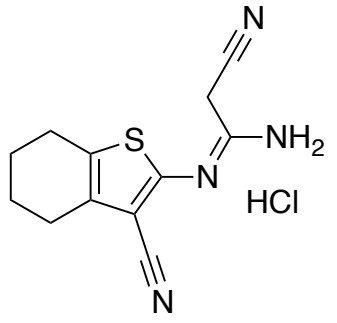 | **SCAM** (0.71) | Non-SCAM | Non-SCAM | Non-SCAM* (0.70) | SCAM |
| B2_A 11 | 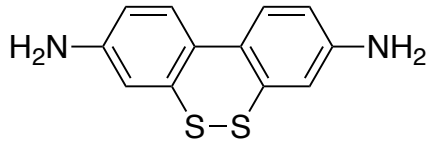 | SCAM (0.69) | **Non-SCAM** | **Non-SCAM** | **Non-SCAM*** (0.80) | Non-SCAM |
| B2_A 12 | 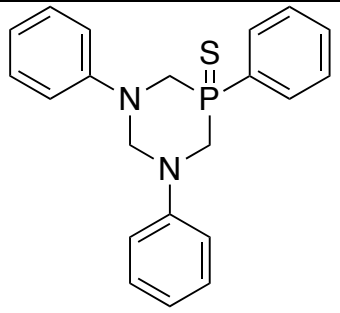 | **SCAM** (0.80) | Non-SCAM | Non-SCAM | Non-SCAM (0.70) | SCAM |
| B2_A 13 | 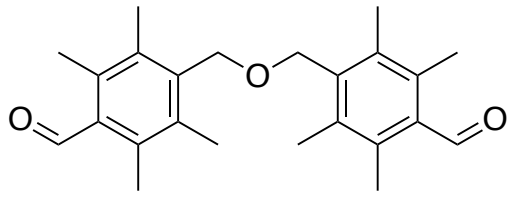 | **SCAM** (0.61) | Non-SCAM | Non-SCAM | Non-SCAM (0.70) | SCAM |
| B2_A 14 | 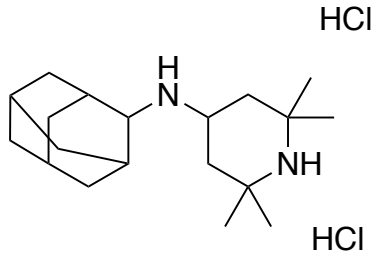 | SCAM (0.76) | **Non-SCAM** | **Non-SCAM** | **Non-SCAM*** (0.70) | Non-SCAM |
| B2_A1 5 | 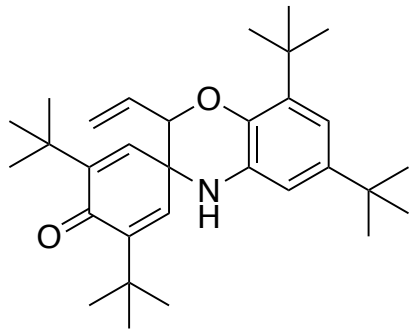 | **SCAM** (0.72) | Non-SCAM | Non-SCAM | Non-SCAM* (0.60) | SCAM |
| B2_A 16 | 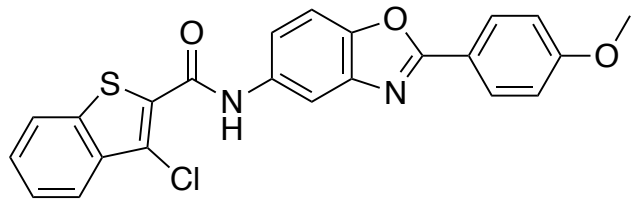 | **SCAM** (0.92) | **SCAM** (0.80) | Non-SCAM | Non-SCAM (0.50) | SCAM |
| B2_A 17 | 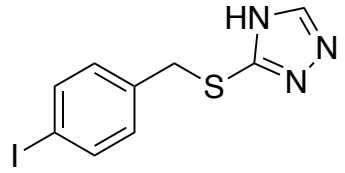 | **SCAM** (0.82) | Non-SCAM | Non-SCAM | Non-SCAM* (0.80) | SCAM |
| B2_A 18 | 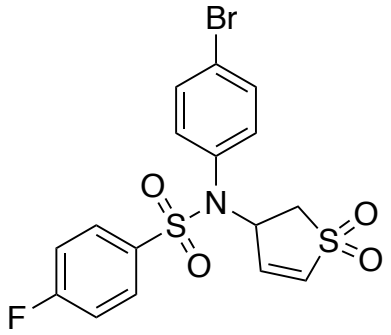 | SCAM (0.93) | **Non-SCAM** | **Non-SCAM** | **Non-SCAM*** (0.70) | Non-SCAM |
| B2_A 19 | 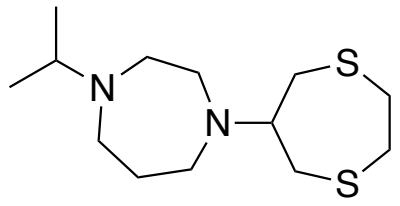 | **Non-SCAM** (0.91) | **Non-SCAM** | **Non-SCAM** | **Non-SCAM*** (0.70) | Non-SCAM |
| B2_A 20 | 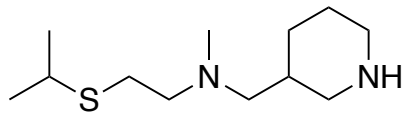 | **SCAM** (0.76) | Non-SCAM | Non-SCAM | Non-SCAM* (0.60) | SCAM |
| Accuracy (%) | | 80[b] | 63[c] | 69[d] | 57[e] | - |
| Precision (%) | | 80[b] | 82[c] | 82[d] | 71[e] | - |
| Recall (%) | | 77[b] | 29[c] | 45[d] | 16[e] | - |

[a]Tanimoto index relative to the nearest neighbour. [b]Size of the training data is 916 molecules. [c]Size of the reference set is >12,500 molecules. [d]Size of the training data is 36,291 molecules. [e]Size of the training data is 272,611 molecules (β-lactamase data). Bold font and green background denote correct predictions. *Outside domain of applicability.

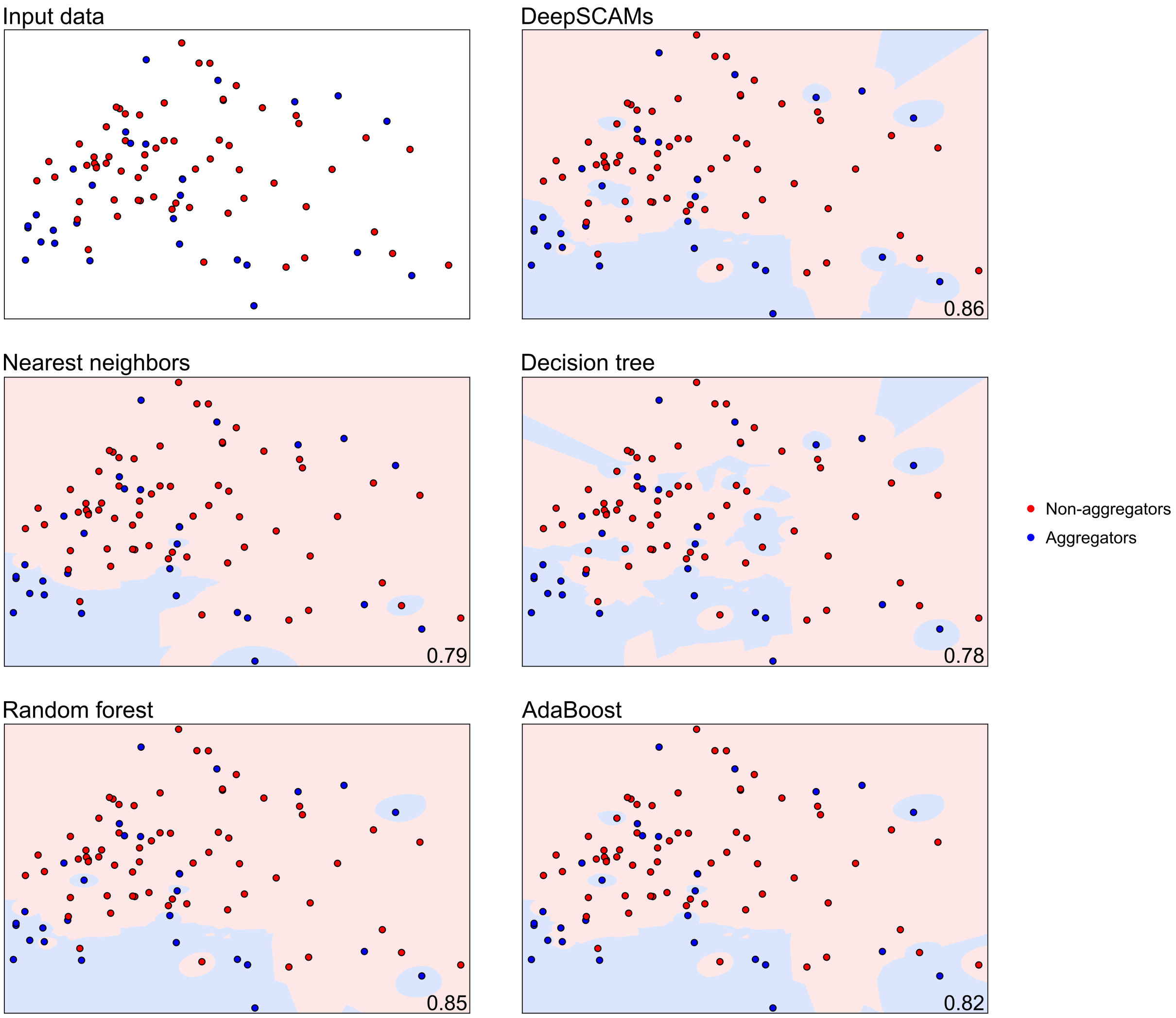

**Figure S1.** Performance of DeepSCAMs relative to myriad computational methods. a) Input data without heuristicaly-assigned class. b) DeepSCAMs (accuracy = 86%). c) Nearest neightbours (accuracy = 77%); d) Decision tree (accuracy = 82%). e) Random forest (accuracy = 89%). f) AdaBoost (accuracy = 80%). Data supports the superior performance of DeepSCAMs over other methods. Decision boundaries are provided by shaded regions: Red, prediction as non-SCAM; Blue, prediction as SCAM. True (experimentally confirmed) classes are denoted in dot colors: Red, non-SCAMS; Blue, SCAMs. Data was projected by Isomap. As an alternative, we performed a principle component analysis (PCA), yet the first two principle components explained only 3.6 and 3.8% of the data variance, rendering it unsuitable for downstream computations in this particular case.

**Table S2.** Hyperparameters screened in miscellaneous machine learning algorithms used for comparison to DeepSCAMs.

| **Model** | **Hyperparameters** | **Value** | **Optimized value** |
|---|---|---|---|
| *k* Nearest neighbors | n_neighbors | 1, 3, 5, 7, 9, 11, 13, 15 | 11 |
| | weights | uniform, distance | uniform |
| | metric | Euclidean, Manhattan | Euclidean |
| Decision tree | criterion | gini, entropy | entropy |
| | splitter | best, random | random |
| | max_depth | 1, 3, 5, 7, 9, 11, 13, 15 | 11 |
| Random forest | n_estimators | 1, 10, 100, 1000 | 100 |
| | criterion | gini, entropy | entropy |
| | max_features | auto, sqrt, log2 | auto |
| | max_depth | 1, 3, 5, 7, 9, 11, 13, 15 | 11 |
| AdaBoost | base_estimator | decision tree<br>max_depth = 1, 5, 10, 15 | decision tree<br>max_depth = 5 |
| | n_estimators | 1, 10, 100, 1000 | 1000 |

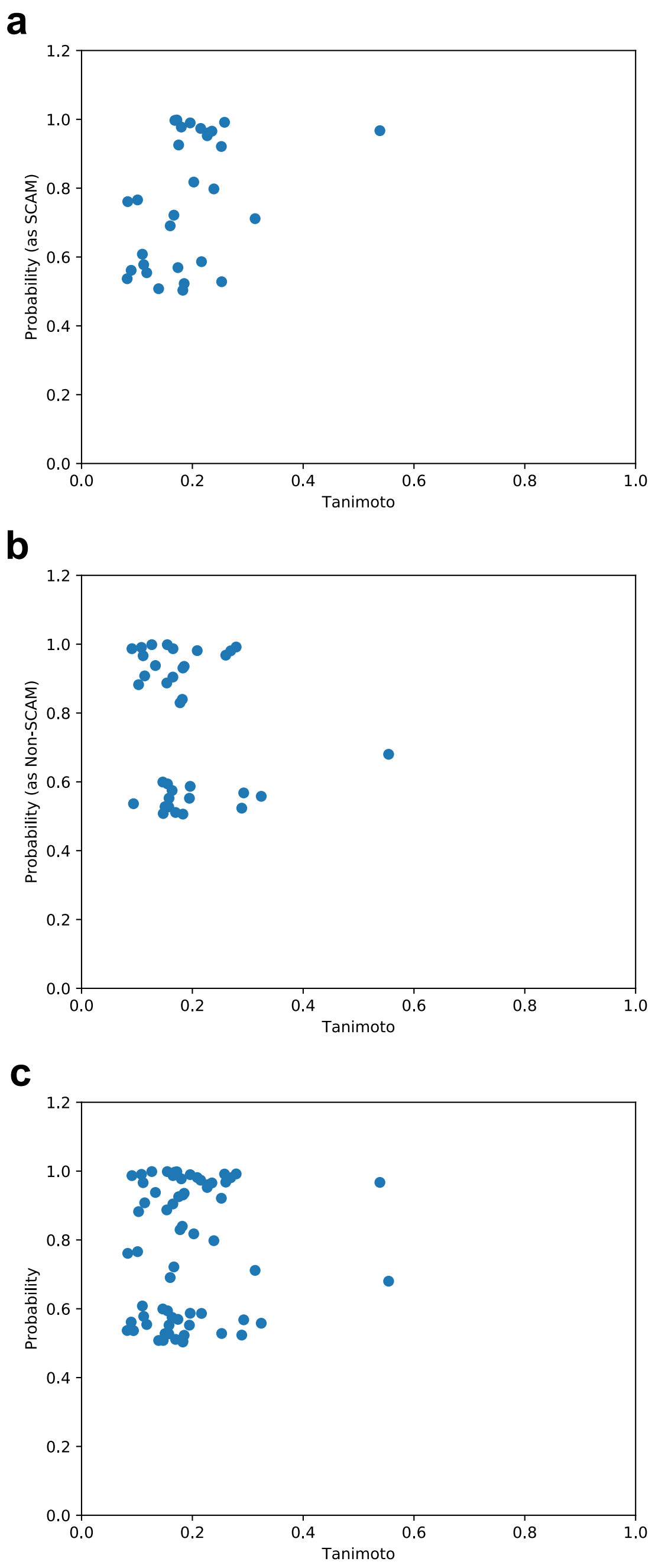

**Figure S2.** Correlation between the class probability for each of the 65 molecules in the prospective validation set and the Tanimoto value relative to the nearest neighbour in the DeepSCAMs training set. a) Correlation for molecules predicted as SCAMs (Pearson $r$ = 0.414; $p$ = 0.02). b) Correlation for molecules predicted as non-SCAMs (Pearson $r$ = 0.159; $p$ = 0.362). Data shows an absence of correlation between substructural similarity to the nearest neighbour and the predicted output, supporting that DeepSCAMs learned relevant data patterns. c) Correlation with the predicted class probability (Pearson $r$ = 0.097; $p$ = 0.44). Data shows an absence of correlation between substructural similarity to the nearest neighbour and the predicted output.

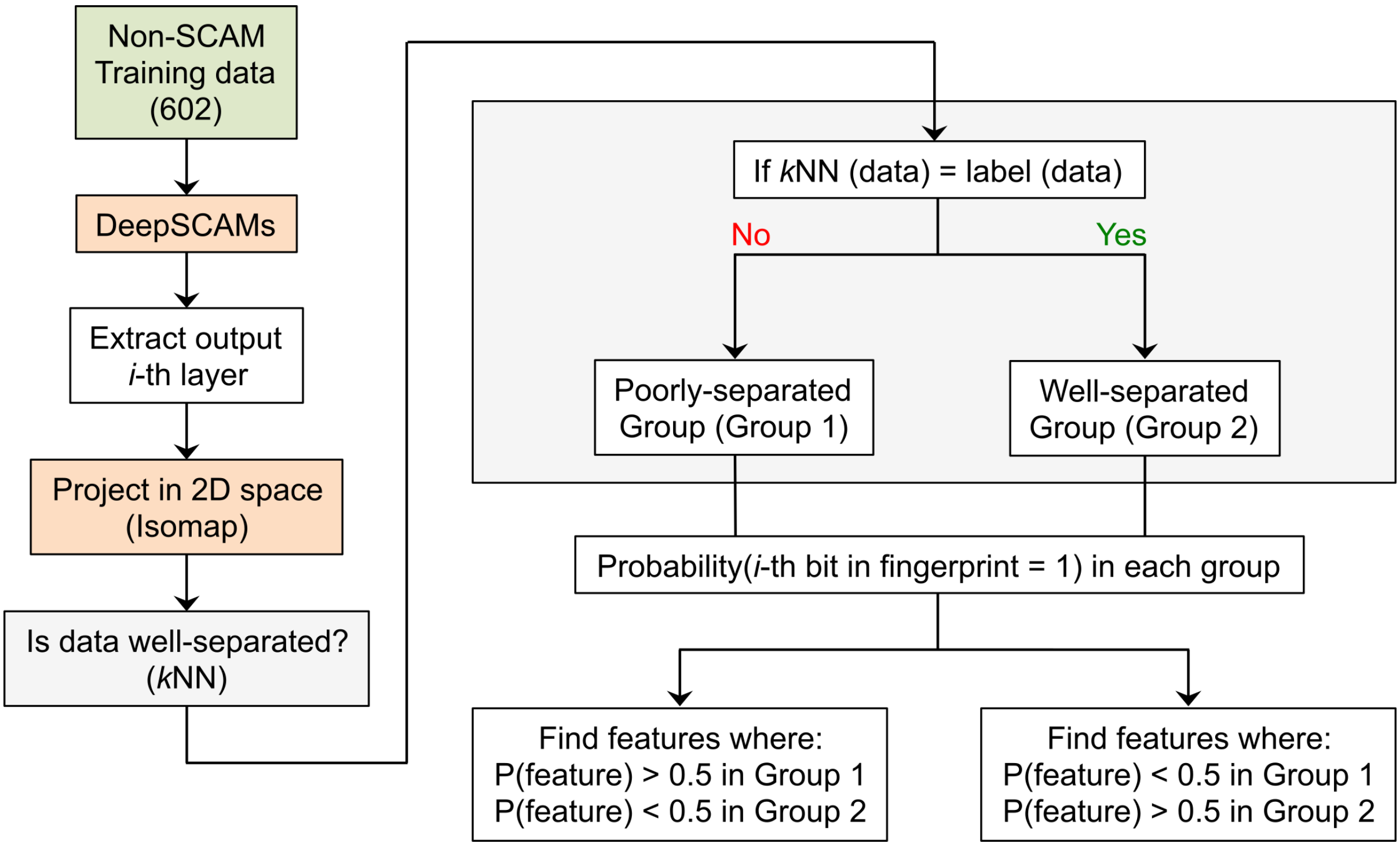

**Figure S3.** Workflow for the identification of well- and poorly separated data.

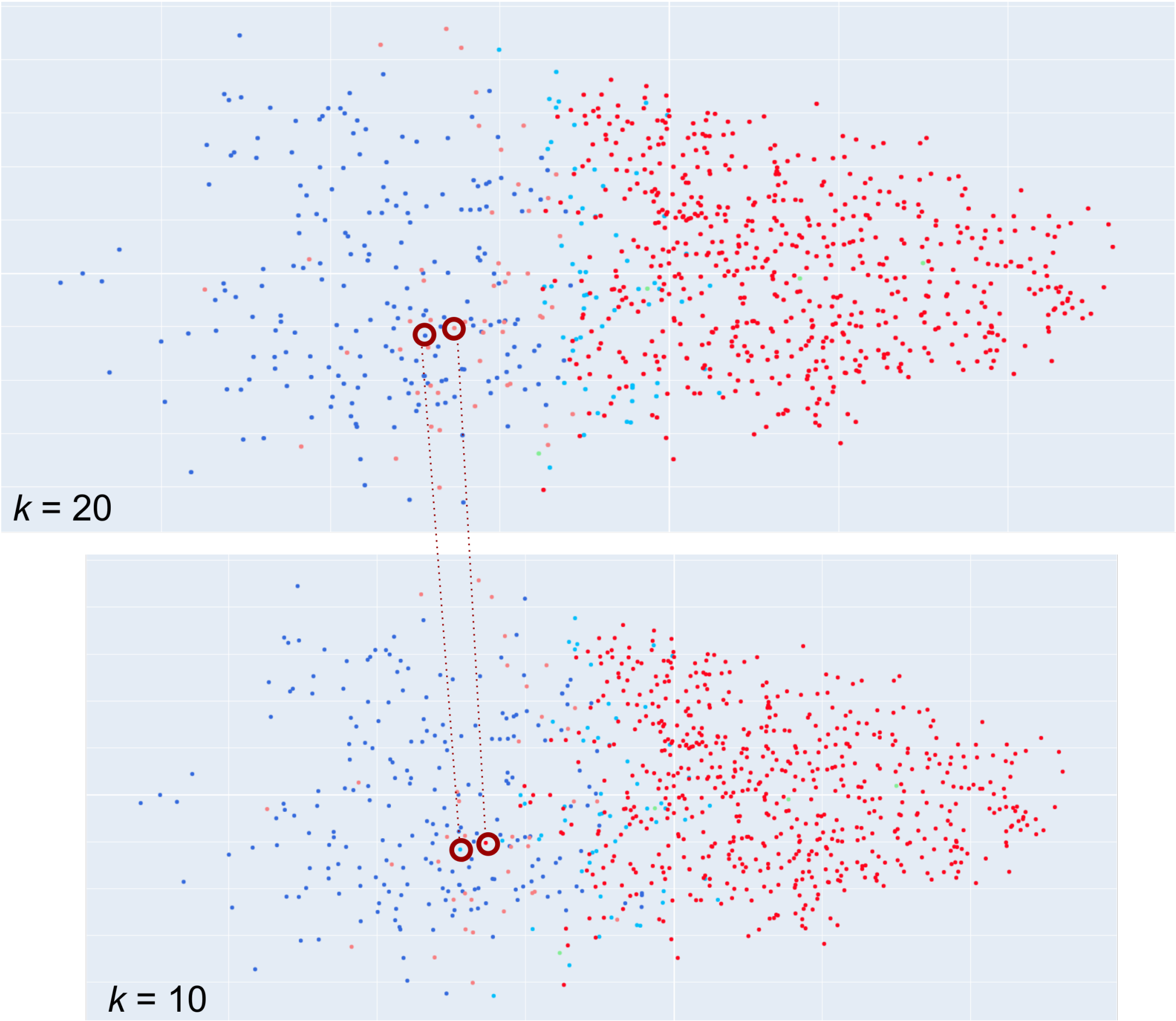

**Figure S4.** Justification of well- and poorly-separated data in Isomaps. The definition between well and poorly-separated data is provided by feeding the 2D data in Isomaps to a k-nearest neighbor algorithm (*k* = 10 or 20). A *k* value of 20 provided better data resolution and was used for all subsequent analyses. For example, if most of a given SCAM neighbors' are non-SCAMs then the SCAMs is poorly-separated, irrespective of being correctly of incorrectly calssified by DeepSCAMs. This allowed the identification of common features in poorly-separated data in poorly-separated but not in well-separed data, and vice-versa. A common feature in a group of data is, therefore, the probability of occurrence of the feature in this group of data, *i.e.* a probability >50% means that a given feature is shared among the group. Dot colors relate to true (experimentally confirmed) classes. Red: non-SCAM; Blue: SCAM. Dark red and blue correspond to well-separed data whereas light red and blue correspond to poorly-separated data.

**Table S3.** Summary of substructural fingerprint bits correlated with well- and poorly-separated data.

| | | Non-SCAM | | SCAM | |
|---|---|---|---|---|---|
| **ANN layer** | | Well-separated | Poorly-separated | Well-separated | Poorly-separated |
| 1 | No. of data | 602 | 50 | 203 | 60 |
| | Common bits exclusive in group | 1152, 1057, 935, 80, 1917 | 875, 694, 1199 | 875 | 80, 1057, 695 |
| 2 | No. of data | 634 | 19 | 238 | 25 |
| | Common bits exclusive in group | 80, 1057, 935 | 675, 725, 694, 1199 | 1199 | 80, 695 |
| 3 | No. of data | 643 | 10 | 256 | 7 |
| | Common bits exclusive in group | 1057, 935 | 675, 725, 694, 1199 | 115, 875 | 314, 389, 935 |

Data suggests that bits 875 and 1199 are important for obtaining well-separated SCAMs, whereas bits 80, 935 and 1057 are important for obtaining well-separated non-SCAMs.

**Table S4.** Significance of select features for distinguishing labels and groups, based on layer 3 data.

| Label | Group | Number of experiments | % Success estimate (probability ± s.d.) |
|---|---|---|---|
| SCAM ($n$ = 263) | Match with poor/well separated ($n$ = 256) | 100 | 0 ± 0 |
| | Match only with poor separated ($n$ = 7) | 100 | $6\times10^{-3} \pm 2\times10^{-4}$ |
| Non-SCAM ($n$ = 653) | Match with poor/well separated ($n$ = 643) | 100 | $3\times10^{-7} \pm 2\times10^{-6}$ |
| | Match only with poor separated ($n$ = 10) | 100 | $6\times10^{-6} \pm 8\times10^{-6}$ |

Note: each experiment contains 100,000 random trials and % success estimate is calculated as a ratio between the number of successes and the total number of trials.

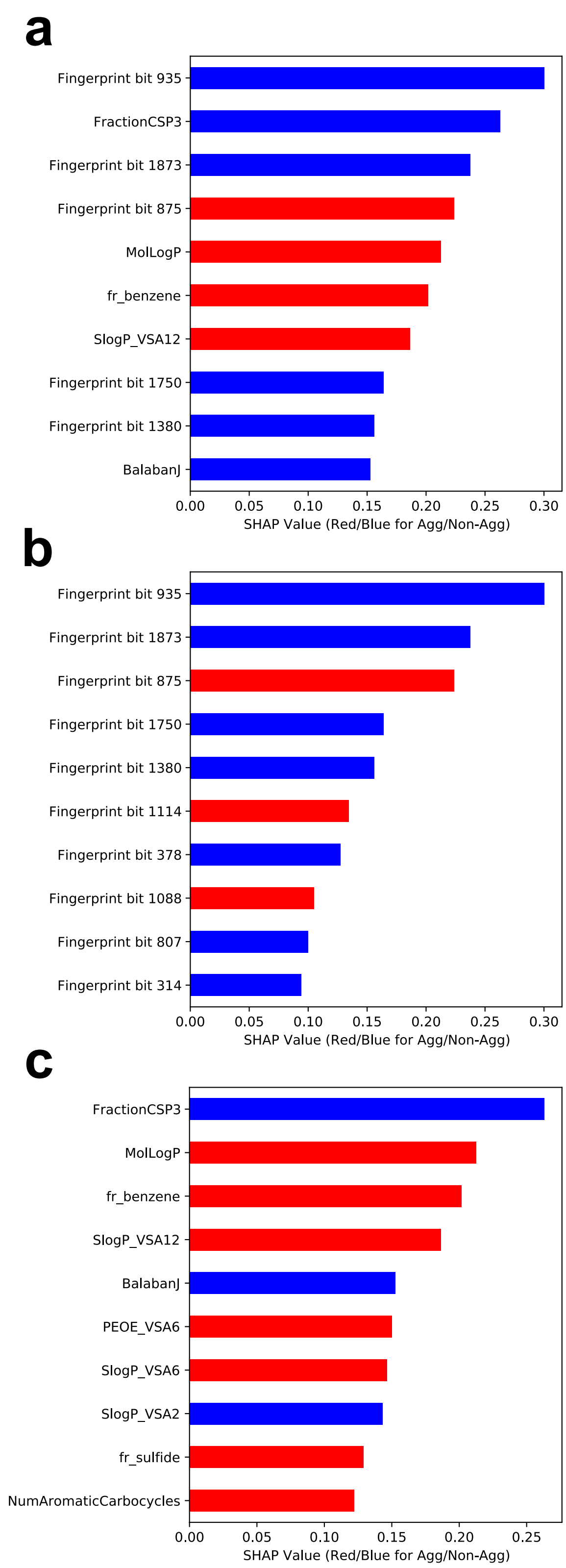

**Figure S5.** Interpretation of predictions for test molecules using SHAP. a) Global feature importance. b) Importance of Morgan fingerprint bits (2048, radius 3). c) Importance of physicochemical features. Red: support aggregation. Blue: support non-aggregation.

**a**

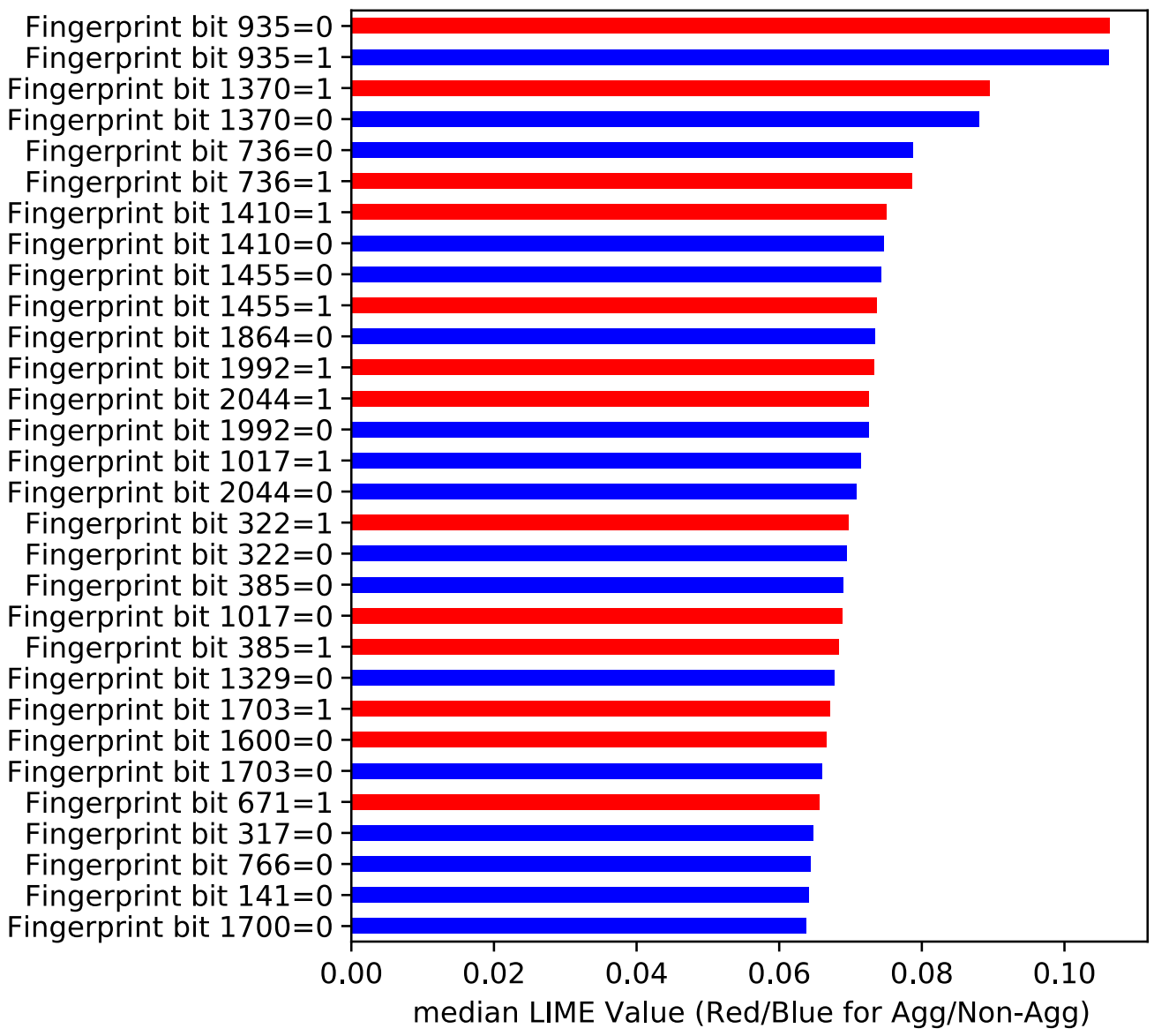

**b**

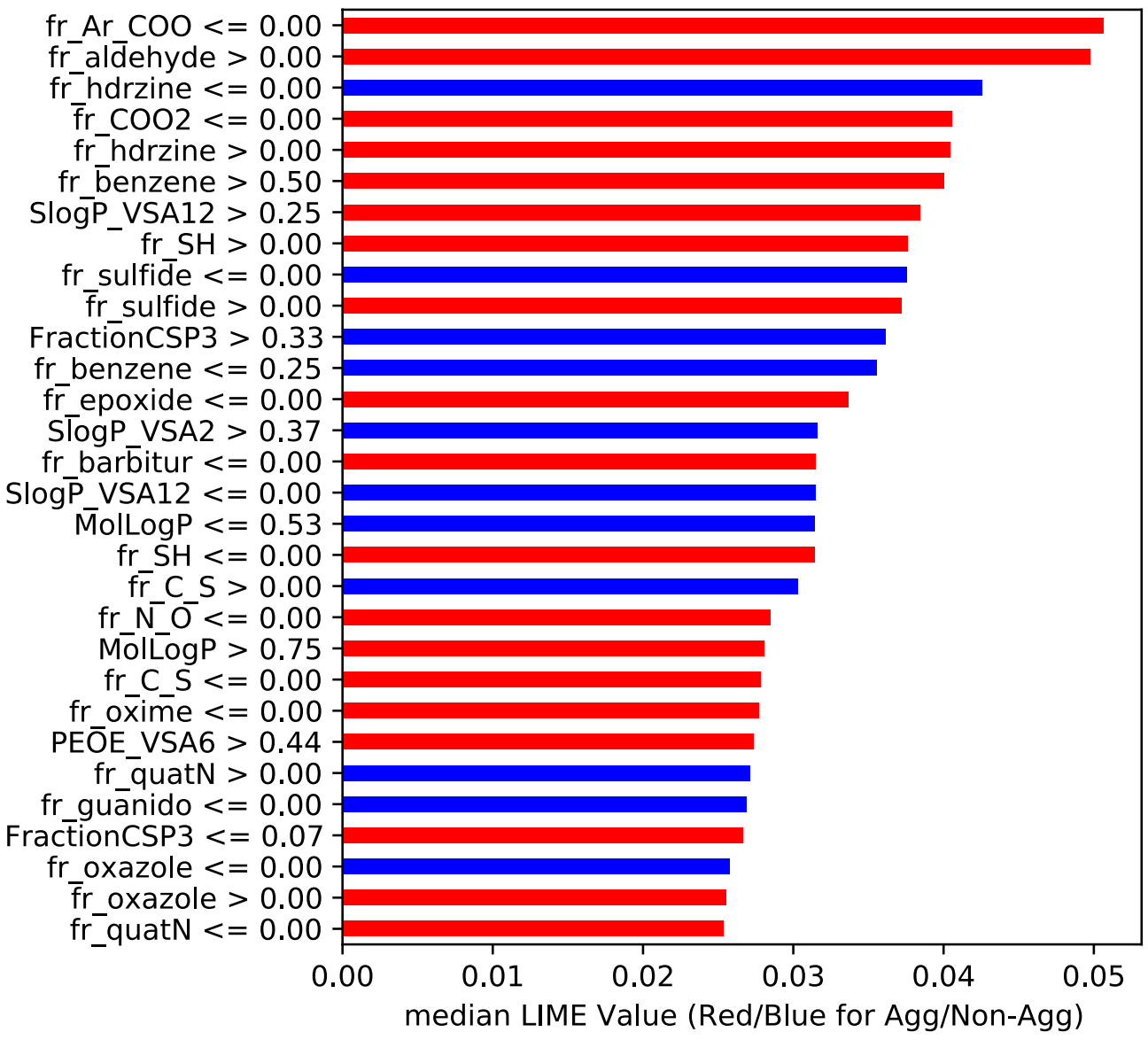

**Figure S6.** Interpretation of predictions for test molecules using LIME. a) Importance of Morgan fingerprint bits (2048, radius 3). b) Importance of physicochemical features. Red: support aggregation. Blue: support non-aggregation.

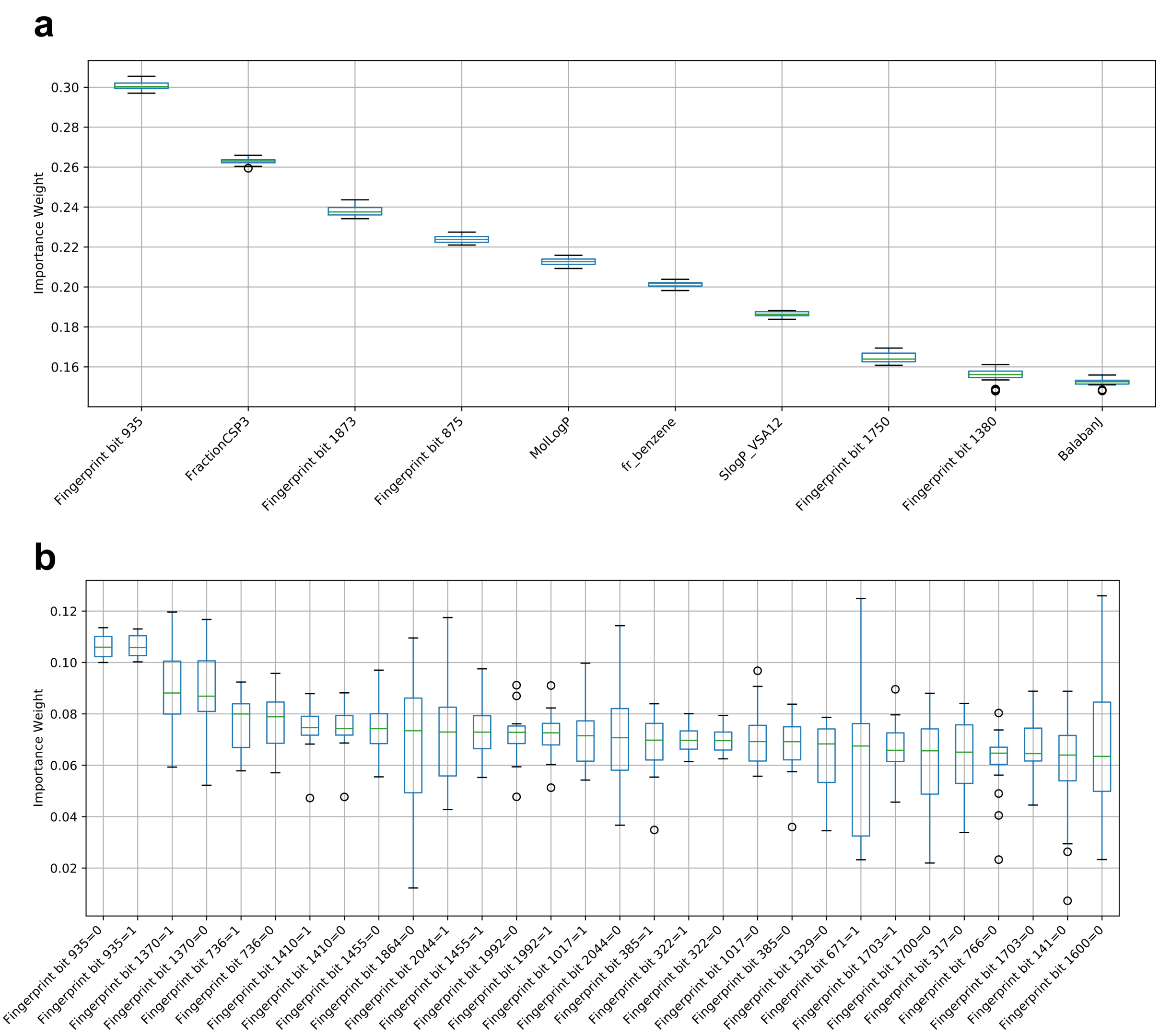

**Figure S7.** Feature importance for test set according to different analysis methods. a) SHAP. Data suggests the high importance of fingerprint bit 935 for the class prediction as well as structural and physicochemical properties (fraction of $sp^3$ carbon atoms, calculated log*P* and fraction of phenyl rings. b) LIME. Data confidently supports the importance of fingerprint bit 935 (low variance).

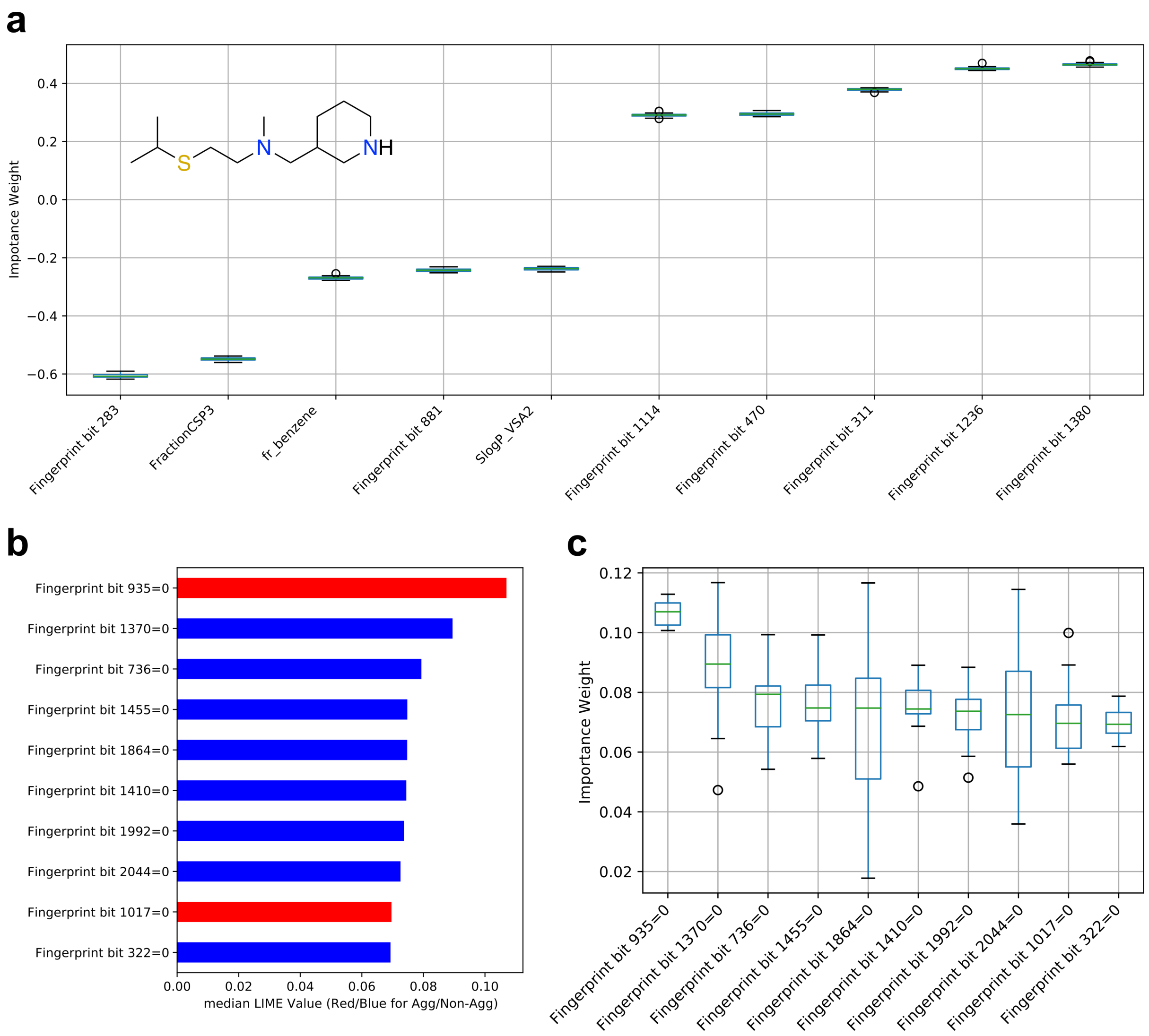

**Figure S8.** Interpretation and robustness of the predicted class for the depicted compound using two methods. a) SHAP values suggest the importance of fingerprints bits 1380, 1236, 311, 470 and 1114. b) Median Lime values for the most important features supporting either aggregation or non-aggregation. c) Variability of the LIME values for each of the most important features.

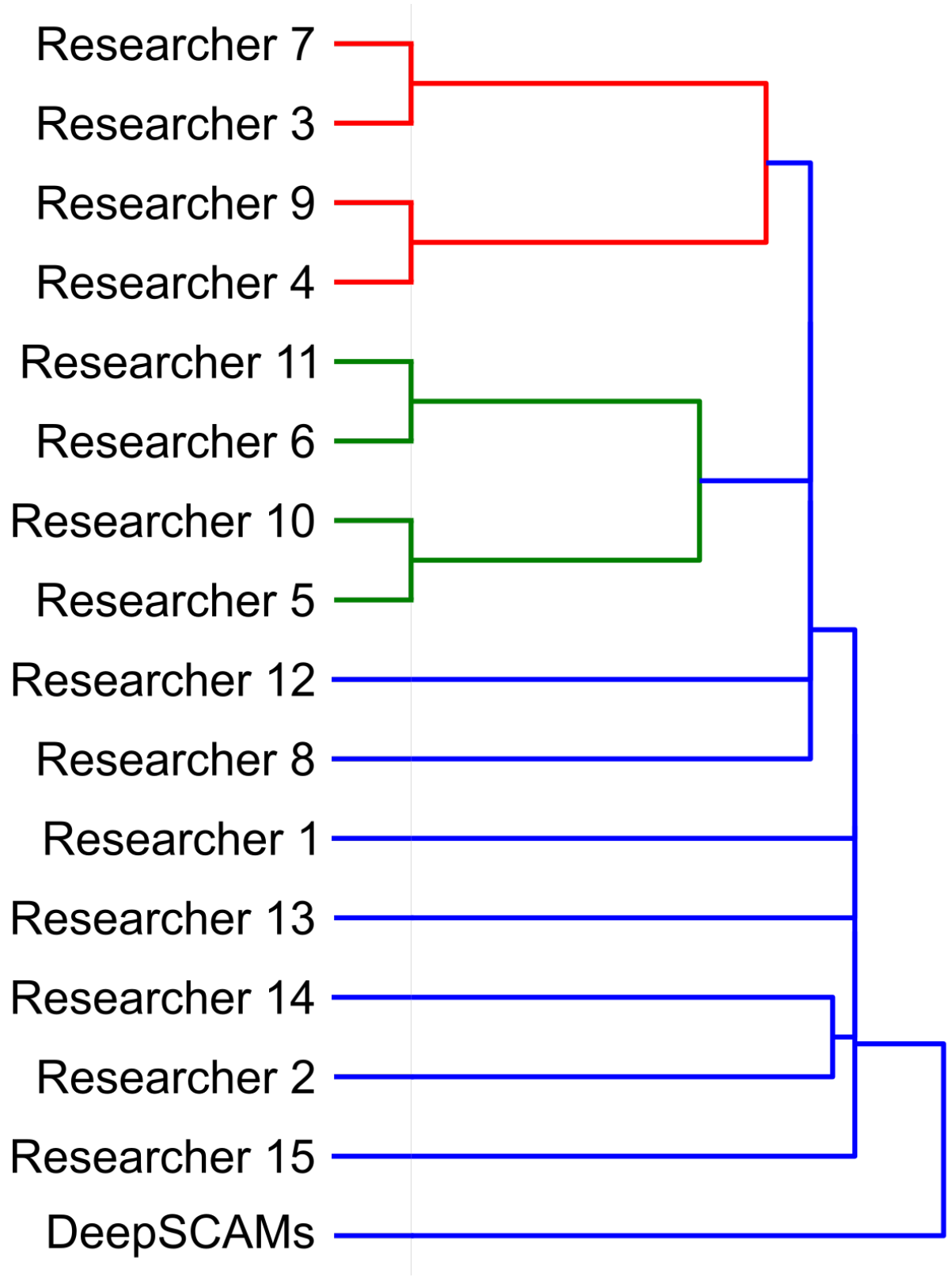

**Figure S9.** Hierarchical clustering of expert and DeepSCAMs 'SCAM'/'non-SCAM' responses for the prospective test set. Data shows the dissimilarity between the automated workflow and responses obtained by expert chemists (similarity metric: Hamming).

**Note S1.**

Both of SHAP and LIME explain the model prediction by training another interpretation model locally around given predictions. The interpretation provides a set of sparse, important weights that instantiate focal points on the model for its decision. Since both of methods are based on re-training linear explanatory models on high dimensional descriptor space, one might expect potential high variance during model fitting. Therefore, we provide variance estimation by repeated experiments (20 times). The results indicate that SHAP produces much more robust explanations than LIME (Supplementary Figure S7). Note that LIME binarizes the original data. Therefore, each fingerprint descriptor will be present twice, similar to one-hot encoding. For example, there will be both zero (absent) and one (present) for each fingerprint bits. We depicted the top-30 features (Supplementary Figure S7). The fingerprint bit 935 is more consistently identified by both models.